\documentclass{jpp}

\usepackage{hyperref}
\usepackage{graphicx}

\usepackage{mathtools}
\usepackage{physics}

\usepackage[utf8]{inputenc}
\usepackage[T1]{fontenc}
\usepackage{amsmath}

\usepackage[dvipsnames]{xcolor}
\usepackage[english]{babel}
\usepackage[autostyle]{csquotes}
\usepackage{float}
\usepackage{placeins}

\usepackage{tikz}
\usepackage{circuitikz}
\usetikzlibrary{calc}
\usetikzlibrary{patterns}
\usetikzlibrary{arrows}
\usetikzlibrary{shapes}
\usetikzlibrary{plotmarks}
\usetikzlibrary{decorations.markings}

\usepackage{pgfplots}
\usepgfplotslibrary{polar}
\usepgflibrary{shapes.geometric}
\usepackage[]{pstricks}

\usepackage{verbatim}
\usepackage{amsfonts}

\usetikzlibrary{shapes.geometric, arrows,chains}
\usetikzlibrary{shapes.geometric, arrows}
\usetikzlibrary{decorations.pathreplacing,calligraphy}
\usetikzlibrary{quotes,angles}

\usepackage{multirow}
\usepackage{threeparttable}
\usepackage{threeparttablex}
\usepackage{tabularx}
\usepackage{tabulary}
\usepackage{array}
\usepackage{multicol}
\usepackage{booktabs}

\newcommand{\rmd}{\mathrm{d}}
\renewcommand{\vec}{\boldsymbol}

\newcommand \s {\nu}
\newcommand \gyror {\boldsymbol{R}_{\s}}

\newcommand \gks {\hat{g}_{\boldsymbol{k} , \s} }
\newcommand \pk {\hat{\phi}_{\boldsymbol{k}} }
\newcommand \ak {\hat{A}_{\parallel, \boldsymbol{k}} }
\newcommand \bk {\delta \hat{B}_{\parallel, \boldsymbol{k}} }

\newcommand \gksz {\bar{g}_{\boldsymbol{k} , \s, 0} }
\newcommand \chikz {\bar{\chi}_{\boldsymbol{k}, 0} }
\newcommand \pkz {\bar{\phi}_{\boldsymbol{k}, 0} }
\newcommand \akz {\bar{A}_{\parallel, \boldsymbol{k}, 0} }
\newcommand \bkz {\delta \bar{B}_{\parallel, \boldsymbol{k}, 0} }

\newcommand \gun {g_{\s}}
\newcommand \pun {\phi}
\newcommand \aun {A_{\parallel}}
\newcommand \bun {\delta B_{\parallel}}

\newcommand \bdotgradzun {\hat{\boldsymbol{b}}\cdot \grad z}

\newcommand \der [2]{\ensuremath {\frac{\rmd #1}{\rmd #2}}}
\newcommand \gradpar {\tilde{v}_{th,\s} \tilde{v}_{\parallel} \; \hat{\boldsymbol{b}}\cdot \tilde{\grad} \tilde{z}}

\newcommand \vthnorm {\tilde{v}_{\text{th},\s}}
\newcommand \vthref {v_{\text{th},i}}
\newcommand \vthspec {v_{\text{th},\s}}

\newcommand \collisionop {C^{(l)}_{\s}}
\newcommand \collisionopadj {C^{(l) \dagger}_{\s}}
\newcommand \gyrochi {\hat{\chi}_{\vec{k}, \s} }

\newcommand \dvpacoeff {\tilde{v}_{th,\s} \; \tilde{\mu}_{\s} \; \hat{\boldsymbol{b}} \cdot \tilde{\grad} \tilde{z} \pdv{\tilde{B}_0}{\tilde{z} } }

\newcommand \velspace {\frac{2\tilde{B}}{\sqrt{\pi}}}
\newcommand \gk {\tilde{g}_{\boldsymbol{k}, \s}}

\newcommand \phick {\tilde{\phi}_{\boldsymbol{k}, 0}}
\newcommand \gck {\tilde{g}_{\boldsymbol{k},\s, 0}}
\newcommand \zt {\frac{Z_{\s}}{\tilde{T}_{\s}}}
\newcommand \besselk {J_{0,\boldsymbol{k}, \s}}
\newcommand \maxwellian {F_{0,\s}}
\newcommand \maxwexp {e^{-\tilde{v}_{\s}^2}}

\newcommand \lamspec {\lambda_{\s}}

\newcommand \bessel {J_{0,\s}}

\newcommand \bp {\boldsymbol{p}}
\newcommand \gspec {\tilde{g}_{\s}}
\newcommand \tilphi {\tilde{\phi}}
\newcommand \tila {\tilde{A}_{\parallel}}
\newcommand \tilb {\delta \tilde{B}_{\parallel}}

\newcommand \tildlam {\lambda^{\leftrightarrow}_{\s}}
\newcommand \tildlamconj {\lambda^{\leftrightarrow, *}_{\s} }

\newcommand \steltildlam {\tilde{\lambda}^{\leftrightarrow}_{\s} }

\newcommand \lamstella {\tilde{\lambda}_{\s} }
\newcommand \xistella {\tilde{\xi} }
\newcommand \zetastella {\tilde{\zeta} }
\newcommand \sigstella {\tilde{\sigma} }

\newcommand{\apref}{Appendix~\ref}
\newcommand{\secref}{Section~\ref}

\newcommand \inprodz [2]{\left\langle #1, #2\right\rangle_{z} }
\newcommand \inprodv [2]{\left\langle #1, #2\right\rangle_{v, {\s}} }
\newcommand \inprodzv [2]{\left\langle #1, #2\right\rangle_{z, v, {\s}} }

\newcommand \gradp {\grad_{\bp}}
\hypersetup{
  colorlinks,
  linkcolor={blue!80!black},
  citecolor={blue!80!black},
  urlcolor={blue!80!black},
}

\title{Optimisation of Gyrokinetic Microstability Using Adjoint Methods}
\author{G. O. Acton \aff{1,2}, M. Barnes \aff{1}, S. Newton \aff{2}, H.~Thienpondt\aff{3}}

\affiliation{\aff{1} Rudolf Peierls Centre For Theoretical Physics, University of Oxford, Oxford, OX1 3PU, UK
\aff{2} United Kingdom Atomic Energy Authority, Culham Campus, Abingdon, Oxfordshire, OX14 3DB, UK
\aff{3} Laboratorio Nacional de Fusi\'on, CIEMAT, 28040 Madrid, Spain }

\begin{document}

\maketitle

\begin{abstract}
Microinstabilities drive turbulent fluctuations in inhomogeneous, magnetized plasmas. In the context of magnetic confinement fusion devices, this leads to an enhanced transport of particles, momentum, and energy, thereby degrading confinement. In this work, we elaborate on the application of the adjoint method to efficiently determine the variation of linear growth rates for plasma microstabilities concerning a general set of external parameters within the local $\delta \! f$-gyrokinetic model. We then offer numerical verification of this approach. When coupled with gradient-based techniques, this methodology can facilitate the optimization process for the microstability of the confined plasmas across a high-dimensional parameter space. We present a numerical demonstration wherein the ion-temperature gradient (ITG) instability growth rate in a tokamak plasma is minimized with respect to flux surface shaping parameters. The adjoint method approach demonstrates a significant computational speed-up compared to a finite-difference gradient calculation.


\end{abstract}
\section{Introduction} \label{sec:intro}

Within the plasma core of magnetic confinement fusion (MCF) devices, gradients of temperature and density act as sources of free energy that are capable of driving microscale instabilities, should they exceed critical values (see e.g. \citealt{Rudakov61}, \citealt{drake1977kinetic}, or \citealt{tang1980kinetic} for ITG, microtearing modes (MTM) or kinetic ballooning modes (KBM) descriptions respectively). Beyond such critical values the transport fluxes increase rapidly, thus requiring a large additional power input to maintain the temperature gradient. 
One mechanism behind these large fluxes is turbulent mixing, which results in the increased transport of particles, momentum, and energy out of the device. To properly predict plasma profiles, one should thus solve the transport equations in which nonlinear, turbulent fluxes determine profile evolution.
However, due to this \enquote*{stiffness} of the transport \citep{Dimits00}, the critical gradients obtained from a linear instability analysis are often a first reasonable approximation to the experimental outcomes
\footnote{An important caveat to mention is that true critical gradient values can be shifted by nonlinear effects. Linear instabilities grow until nonlinear coupling to \enquote*{zonal} modes removes energy from the linearly growing modes. This energy injection into the zonal modes causes them to grow and can suppress turbulence completely in a region beyond the linear critical gradients (e.g. see \citealt{Dimits00}, and \citealt{Rogers00}). We will neglect this \enquote{Dimits shift} of the critical gradient in the following analysis, and use only the linear gradient as a predictor for experimental profiles. This provides a lower bound for critical gradient values and can be a good approximation for the full nonlinear threshold value.}.
Hence, it is worthwhile to determine if externally-controlled experimental parameters can be chosen to optimise linear stability.

The growth rate of linear microinstabilities can be influenced by a large number of parameters, including plasma density, temperature, flow profiles, and the magnetic geometry. In an idealised situation, one would determine how such growth rates, and by extension critical gradients, depend on all parameters governing the system, and then design MCF devices that maximise temperature and/or density gradients. 
However, because the number of tunable parameters in modern MCF devices is large, analytical searches are intractable, and full numerical parameter scans are prohibitively expensive.  

In this paper we address this issue by employing an adjoint approach (cf. \citealt{Pironneau74}) to enable efficient calculation of local variations to the linear growth rate with respect to all parameters of interest within the local, $\delta \! f$-gyrokinetic model. In contrast to a finite-difference calculation, the adjoint method is essentially independent of the dimension of the parameter space, and can be used to optimise over a large number of variables at once without incurring additional computational cost beyond solving the system equations; the associated cost of the adjoint method is roughly equivalent to solving the linearised gyrokinetic system of equations twice. 


The archetypal microinstability in MCF plasmas is the ion temperature gradient (ITG) instability (see e.g. \citealt{Rudakov61}, and \citealt{Cowley91}), as it has been identified as the main source of heat transport in the core of many tokamaks. Therefore, for clarity, we shall present the optimisation process with ITG instabilities in mind. We shall apply the adjoint method to the linear gyrokinetic equation and demonstrate its utility by calculating the sensitivity of the growth rate to geometrical parameters in a tokamak, with the aim of maximising the critical ion temperature gradient through shaping considerations only. However, the technique and equations derived below allow for gradient calculations of the linear growth rate with respect to a general set of unspecified parameters, $\bp$, and could thus be applied more broadly.

This paper is organised as follows: in  \secref{sec:plasma_evolution} we outline the gyrokinetic-Maxwell system, including the governing equations for the evolution of the plasma distribution function and electromagnetic fields. 
In \secref{sec:Adjoint_method_GK}, we present details of the gyrokinetic-adjoint system, and derive the adjoint equations as applied to a linear, low-flow, electromagnetic $\delta \! f$-gyrokinetic system with collisions. The aim in mind is to minimise the linear growth rate in a parameter space defined by a generic set of independent variables, $\bp$, that influence the plasma evolution. We also extract the electrostatic, collisionless limit of the adjoint equations. The second portion of the paper then focuses on applying the model derived, and discussing how this method may be used to vary the plasma profile (such as temperature or density gradients) in directions favourable for fusion whilst retaining microstability. 
In \secref{sec:implementation_into_code} we briefly discuss how the process is implemented using the  $\delta \! f$-gyrokinetic code \texttt{stella}, with \secref{sec:adjoint_miller} specialising to the case where $\vec{p}$ consists of the parameters needed to specify the local magnetic geometry within the Miller formalism \citep{Miller98}; we describe the motivation behind this choice, and describe how it is implemented numerically. 
Section \ref{sec:numerical_implementation} covers the numerical methods employed in solving the adjoint system. Finally, Section \ref{sec:Numerical_results} presents the numerical results, including tests for an example application in which the adjoint method is utilised to vary the magnetic geometry with the aim to minimise the linear ITG growth rate.


\section{Plasma Evolution Equations}\label{sec:plasma_evolution}
\noindent
We model the evolution of plasma fluctuations with the $\delta \! f$-gyrokinetic equation (\cite{catto1978linearized}; \cite{antonsen1980kinetic}; \cite{Abel13}), which one derives by exploiting spatial and temporal scale separation. Our starting point is the Vlasov-Maxwell system of equations including collisions:


\begin{align}
	\frac{\rmd f_{\s}}{\rmd t} = \sum_{\s'} C_{\s \s'} [f_{\s}, f_{\s'} ] \; , 
	\label{eq:vlasov_maxwell_eqn} \\
	\grad \cdot \vec{E} = 4 \pi \varrho , 
	\label{eq:gauss_law_max1}\\
	\grad \cdot  \vec{B} = 0 , 
	\label{eq:div_b_max1}\\
	\frac{1}{c} \pdv{\vec{B}}{t} = - \curl{\vec{E}} , 
	\label{eq:faradays_law_max1}\\
	\curl{\vec{B}} = \frac{1}{c} \left( 4 \pi  \vec{j} + \pdv{\vec{E}}{t} \right) ,
	\label{eq:amperes_law_max1}
\end{align}
\noindent
where $\varrho$ is the electric charge density, $\vec{j}$ is the current density, $\s$ is the particle species index, $c$ the speed of light, $t$ the time, $ \vec{E} $ and $ \vec{B} $ are the electric and magnetic fields respectively, and $ C_{\s \s'} [f_{\s}, f_{\s'} ] $ accounts for the effects on species $\s$ from Coulomb collisions with species $\s'$. We relate the charge density and current density to the particle distribution function, $ f_{\s} $, via the velocity space integrals

\begin{align}
	\varrho = \sum_{\s} Z_{\s} e \int \rmd^3 v \; f_{\s} , \\
	\vec{j} = \sum_{\s} Z_{\s} e \int \rmd^3 v \; \vec{v} f_{\s}, 
\end{align}

\noindent
with $Z_{\s} e$ the species charge, and $e$ the proton charge.
We take the frequency of fluctuations, $\omega$, to be much less than the gyrofrequency of particles, $\Omega_{\s}$, defined as $\Omega_{\s} = Z_{\s} e B/ m_{\s} c$ with $B$ the magnitude of the magnetic field, and $m_{\s}$ the species mass.

We decompose quantities into their mean and fluctuating parts. The mean components determine the evolution of the background plasma, and are found  by averaging over all fluctuations. We represent the average of a given quantity, $h$, over all fluctuations by $\langle h \rangle_{\text{turb} } $ and define this average as

\begin{equation}
	\langle h (t) \rangle_{\text{turb}}  = \frac{1}{T} \int_{t-T/2}^{t+T/2} \rmd t' \; \langle h(t') \rangle_{\perp} ,
	\label{eq:turbavg_def}
\end{equation}
\noindent
where $T$ is some intermediate time shorter than the (transport) timescale associated with mean profile evolution, and longer than timescales associated with typical fluctuations, such that $\omega^{-1} \ll T \ll L/ \vthref$, where $L$ is the system size, and $\vthspec \doteq \sqrt{2 T_{\s}/m_{\s}}$ the species thermal velocity, with $T_{\s}$ the species temperature. Here $\langle \cdot \rangle_{\perp}$ is an appropriately defined spatial average over a length $l$ satisfying $\rho_{\text{ref}} \ll l \ll L$
for a surface perpendicular to the magnetic field \citep{Abel13}. The distribution function is decomposed as $f_{\s} =  F_{\s} + \delta f_{\s}$ with $F_{\s}  = \langle f_{\s} \rangle_{\text{turb}} $ determining the profile of the equilibrium plasma, and $\delta f_{\s}$ the contribution from plasma fluctuations.  

We take the Larmor radius of particles, $\rho_{\s} = v_{\perp}/\Omega_{\s}$, to be much smaller than the system size, $ L $, where $v_{\perp}$ is the magnitude of velocity in the plane perpendicular to the magnetic field. 
We henceforth take the reference scale to be ordered the same as the ion scale, with $\rho_{\text{ref}} = v_{\text{th}, \text{ref}} /\Omega_{\text{ref}} \sim \rho_i $. We expand the Vlasov-Maxwell equations including collisions, \eqref{eq:vlasov_maxwell_eqn}-\eqref{eq:amperes_law_max1},  in the small parameter $\epsilon \sim \rho_{\star} \ll 1$, with $\rho_{\star} = \rho_{i} / L $ the ratio of the reference ion gyroradius to the system size, and equate terms of equivalent order. We order terms as follows
\begin{align}
	\epsilon \sim \rho_{\star} \doteq \frac{\rho_{i}}{L} \sim \frac{\omega}{\Omega_{i}} \sim \frac{k_{\parallel} }{k_{\perp}} \sim \frac{e \phi}{T_{\s}} \sim \frac{\delta B}{B_0}  \sim \frac{\delta f_{\s}} {\maxwellian} \ll 1 
	\qquad \text{and} \qquad 
	k_{\perp} \rho_{i} \sim 1 \; .
	\label{eq:orderings}
\end{align}

\noindent
In the above orderings we have decomposed the magnetic field into the equilibrium contribution, $\vec{B}_0$, and the magnetic fluctuations, $\delta \vec{B}$, such that the full magnetic field is $\vec{B} = \vec{B}_0 + \delta \vec{B}$. We have also introduced the parallel and perpendicular wavenumbers, $k_{\parallel} = \boldsymbol{k} \cdot \hat{\boldsymbol{b}}$ and $\boldsymbol{k}_{\perp} = (\boldsymbol{I} - \hat{\boldsymbol{b}} \hat{\boldsymbol{b}}) \cdot \boldsymbol{k}$, with $\hat{\boldsymbol{b}}$ the unit vector in the direction of the magnetic field, and $\boldsymbol{I}$ the identity matrix. The perturbed electric potential has been introduced as $\phi$. We also expand the mean and fluctuating components of the distribution function
\begin{align}
	F_{\s} = & F_{0,\s} + F_{1,\s} + F_{2,\s} \cdots , \nonumber \\
	\delta f_{\s} = & \delta f_{1,\s} + \delta f_{2, \s} + \cdots ,
\end{align}
\noindent
with $F_{0,\s} \sim f_{\s}$, $F_{1,\s} \sim \delta f_{1,\s} \sim \epsilon f_{\s} $, $F_{2,\s} \sim \delta f_{2,\s} \sim \epsilon^2 f_{\s}$, and so on. Equilibrium quantities are taken to have characteristic length scales of order $L$, and evolve slowly on the long transport time scale $\tau_{E}^{-1} \sim \epsilon^3 \Omega_{i}$; they are thus understood to be static during our considerations. Small-scale fluctuations, captured in $\delta f_{\s}$, have characteristic length scales of the order $\rho_{i} \sim \epsilon L$ and frequencies $\omega \sim \epsilon \Omega_{i}$. 

The charged particles follow magnetic field lines and perform rapid gyration in the plane perpendicular to the field. We introduce $\gyror$ as the gyro-centre for species $\s$, and $\boldsymbol{r}$ to indicate the spatial position vector for any given particle. These descriptions of particle location are related through $\gyror = \boldsymbol{r} - \boldsymbol{\rho}_{\s} (\vartheta)$, with $\boldsymbol{\rho}_{\s} (\vartheta) = \hat{\boldsymbol{b}}_0 \cross \boldsymbol{v} / \Omega_{\s}$ the velocity-dependent vector gyroradius. Our analysis need only consider $\hat{\vec{b}}_0$; hence we shall set $\hat{\vec{b}} \equiv \hat{\vec{b}}_0$ in the remainder of this paper for notational brevity. The gyrophase, $\vartheta$, characterises this gyro-motion, and has a large associated frequency $|\dot{\vartheta}| \approx \Omega_{\s}$. A natural approach is to average over these fast oscillations by introducing a gyroaverage, defined through
\begin{align}
	\langle h(\vec{r}) \rangle_{\gyror} & = \langle h(\gyror + \vec{\rho}_{\s} (\vartheta)) \rangle_{\gyror} = \frac{1}{2 \pi} \int_0^{2\pi} \; h(\vec{R}_{\s} + \vec{\rho}_{\s} (\vartheta)) \;  \rmd \vartheta ,
	\label{eq:gyroaverage_guiding_centre_constant}
	\\
	\langle h(\gyror) \rangle_{\vec{r}} & = \langle h(\vec{r} - \vec{\rho}_{\s} (\vartheta)) \rangle_{\vec{r}} = \frac{1}{2 \pi} \int_0^{2\pi} \; h(\vec{r} - \vec{\rho}_{\s} (\vartheta)) \; \rmd \vartheta ,
	\label{eq:gyroaverage_particle_position_constant}
\end{align}

\noindent
where $\gyror$ and $\vec{r}$ are held constant when performing the $ \vartheta $ integrations in equations \eqref{eq:gyroaverage_guiding_centre_constant} and \eqref{eq:gyroaverage_particle_position_constant} respectively. 

Expanding the Vlasov-Maxwell equations, then equating terms of order $\epsilon ^{-1}$ with respect to $\vthref F_{0,i} /L$ we obtain the equation $\partial F_{0, \s} / \partial \vartheta = 0$, which demands $F_{0,\s}$ be independent of gyrophase. In the presence of modest collisionality\footnote{This requires the collisionality $\nu_* \gtrsim \rho_* \omega$ -- a regime within which we work.} zeroth order terms provide the further constraint that the equilibrium component is a Maxwellian in velocities:
\begin{equation}
	\maxwellian \coloneqq \frac{n_{\s}}{ (\pi v_{\text{th},\s}^2)^{\frac{3}{2}}} e^{-v^2/v_{\text{th},\s}^2}, 
\end{equation}



\noindent
where $n_{\s}$ represents the species density. 

Equating terms ordered $\epsilon^1$ gives the evolution equation for first order perturbations. This exists within a six dimensional phase space with coordinate choice $\{\gyror, v_{\parallel}, \mu_{\s}, \vartheta\}$. Here $\mu_{\s}$ is the magnetic moment defined as $\mu_{\s} = m_{\s} v_{\perp}^2/ 2 B $ and is a conserved quantity to the order of consideration. Gyrophase dependence is removed by gyroaveraging the full equation, reducing the phase-space dimensionality by one. As a result the gyroaveraged fluctuating distribution function arises, denoted using $g_{\s} (\vec{R}_{\s}, v_{\parallel}, \mu_{\s} ) \doteq \langle \delta f_{\s} \rangle_{\gyror} $, such that $g_{\s}$ is gyrophase-independent.  

We expand the total time derivative in \eqref{eq:vlasov_maxwell_eqn} in terms of partial derivatives in $\{t, v_{\parallel}, \mu_{\s}, \vartheta, \gyror \}$ -- with each partial derivative taken assuming all other variables are held fixed, unless explicitly stated otherwise -- and then gyroaveraging over $\vartheta$ to reduce the dimensionality of our equations from 6 to 5. We define a \enquote*{low-flow}, or \enquote*{drift}, ordering to be when the flow speed is ordered as $\rho_*$ small compared with the thermal speed. In this ordering the resulting linear, electromagnetic gyrokinetic equation including collisions is
\begin{align}
	& \pdv{g_{\s}}{t} + v_{\parallel} \left[ \hat{\boldsymbol{b}} \cdot \grad g_{\s} + \frac{Z_{\s} e}{T_{\s}} \maxwellian \hat{\boldsymbol{b}} \cdot \grad \langle \chi \rangle_{\gyror} \right]
	+ \boldsymbol{v}_{M,\s} \cdot \left[ \grad_{\perp} g_{\s} + \frac{Z_{\s} e}{T_{\s}} \maxwellian \grad_{\perp} \langle \chi \rangle_{\gyror} \right] \nonumber \\ & 
	- \frac{\mu_{\s}}{m_{\s}} \; \hat{\boldsymbol{b}} \cdot \grad B_0 \pdv{g_{\s}}{v_{\parallel}} + \langle \boldsymbol{v}_{\boldsymbol{\chi}} \rangle_{\gyror} \cdot \left. \grad \right|_{\mathcal{E}} \maxwellian
	+ \frac{Z_{\s} e}{T_{\s}} \; \frac{\mu_{\s}}{m_{\s} c } \hat{\boldsymbol{b}} \cdot \grad B_0 \; \maxwellian \langle A_{\parallel} \rangle_{\gyror} 
	= \collisionop \; ,
	\label{eq:full_EM_gyrokinetic_eqn}
\end{align}
\noindent
where $\collisionop = \sum_{\s'} C^{l}_{\s, \s'}$ is the linearised collision operator taken to be self-adjoint such that $\collisionopadj = \collisionop$ \footnote{It should be noted that this encapsulates a broad range of collision operators including the linearised Landau, and the linearised Fokker-Planck collision operators.}. We have introduced $\chi = \phi - \vec{v} \cdot \vec{A}/c$ as the gyrokinetic potential, with $\vec{A} = A_{\parallel} \hat{\vec{b}} + \boldsymbol{A}_{\perp}$ the fluctuating magnetic vector potential, ($\delta \vec{B} = \curl{\vec{A}} $), which has been decomposed into components parallel and perpendicular to the equilibrium magnetic field. The gradient acting on the Maxwellian appears as $\left. \grad \right|_{\mathcal{E}}$. This indicates that the derivative has been taken at constant kinetic energy, $\mathcal{E} = m_{\s} v^2/2$, rather than at fixed $\{v_{\parallel}, \mu_{\s} \}$ variables, in contrast to the other spatial gradients appearing in equation \eqref{eq:full_EM_gyrokinetic_eqn}. Finally, $\boldsymbol{v}_{M,\s}$ and $\boldsymbol{v}_{\vec{\chi}}$ are the magnetic and generalised $\boldsymbol{E} \cross \boldsymbol{B}$ drifts defined through 
\begin{align}
	\boldsymbol{v}_{M,\s} = &  \frac{1}{\Omega_{\s}} \hat{\boldsymbol{b}} \cross \left( \frac{\mu_{\s}}{ m_{\s}}\grad B + v_{\parallel}^2 \boldsymbol{\kappa} \right) , \\
	\boldsymbol{v}_{\chi} = & \frac{c}{B} \hat{\boldsymbol{b}} \cross \grad_{\perp} \chi ,
	\label{eq:untransformed_drifts}
\end{align}
\noindent
with $\boldsymbol{\kappa} = \hat{\boldsymbol{b}} \cdot \grad \hat{\boldsymbol{b}}$ the equilibrium magnetic field curvature. 

The system is closed by the field equations consisting of quasineutrality, ${\sum_{\s} Z_{\s} \delta n_{\s} = 0}$, with $\delta n_{\s}$ the perturbed density, and the low-frequency Amp\`{e}re's law, $ \grad \cross \delta \boldsymbol{B} = (4\pi / c) \delta \boldsymbol{J} $, with $\delta \boldsymbol{B}$ and  $\delta \boldsymbol{J}$ the fluctuating magnetic field and current density respectively. When written in terms of the distribution function these relations become: 
\begin{align}
	\sum_{\s} Z_{\s} \int \rmd^3 v \left[ \langle g_{\s} \rangle_{\gyror} + \frac{Z_{\s} e}{T_{\s}} \maxwellian \left( \left\langle \langle \chi \rangle_{\gyror} \right\rangle_{\vec{r}} - \phi \right) \right]  & = 0, 
	\label{eq:quasineutrality_unnormalised} \\
	\grad_{\perp}^2 A_{\parallel} - \frac{4\pi}{c} \sum_{\s} Z_{\s} e \int \rmd^3 v \; v_{\parallel} \left[ g_{\s} + \frac{Z_{\s}e}{T_{\s} c} \maxwellian \; v_{\parallel} \left\langle \langle A_{\parallel} \rangle_{\gyror} \right\rangle_{\vec{r}} \right] & = 0, 
	\label{eq:par_ampere_law_unnormalised} \\
	\grad_{\perp} \delta B_{\parallel} - \frac{4\pi}{c} \sum_{\s} Z_{\s} e \int \rmd^3 v \; \grad \cdot \left[ \left( g_{\s} + \frac{Z_{\s}e}{T_{\s}} \maxwellian \left\langle \langle \chi \rangle_{\gyror} \right\rangle_{\vec{r}}  \right)(\hat{\boldsymbol{b}} \cross \boldsymbol{v}_{\perp} ) \right] & = 0 ,
	\label{eq:perp_ampere_law_unnormalised}
\end{align}
\noindent
where $\delta B_{\parallel} = \delta \boldsymbol{B}\cdot \hat{\boldsymbol{b}} =  (\grad \cross \vec{A}_{\perp}) \cdot \hat{\boldsymbol{b}}$ is the parallel component of the perturbed magnetic field.


\subsection{Magnetic Coordinates} \label{sec:coordinate_system_stella}
\noindent
We can express the magnetic field using the  Clebsch representation: 

\begin{equation}
	\boldsymbol{B} = \grad \alpha \cross \grad \psi ,
	\label{eq:clebsch_magnetic_field}
\end{equation}
\noindent
where $\psi$ and $\alpha$ are the flux surface and field line labels respectively. We choose to work in field-aligned coordinates \citep{Beer95} denoted by $(x,y,z)$, with $z$ the position along the magnetic field line and $(x,y)$ the position in the plane perpendicular to $\hat{\boldsymbol{b}}$. Together the unit vectors $\{ \hat{\vec{x}}, \hat{\vec{y}}, \hat{\vec{b}}\} $ form a left-handed, orthonormal basis. We relate the coordinates $(x,y)$ to $(\psi, \alpha)$ using

\begin{equation}
	\begin{aligned}
		x = & \frac{d x}{d \psi} (\psi - \psi_0) , \\
		y = & \frac{d y}{d \alpha} (\alpha - \alpha_0) , 
		\label{eq:yx_to_phi_alpha}
	\end{aligned}
\end{equation}

\noindent
with $(\psi_0, \alpha_0)$ denoting the values at the centre of the domain. 

We take the discrete Fourier transform in $\{x,y\}$
whilst retaining the real space coordinate for the parallel direction, and evaluate nonlinear terms pseudo-spectrally in order to retain spectral accuracy in spatial derivatives when implementing the equations numerically\footnote{Spatial derivatives are exact in the Fourier representation and are hence more accurate than using a finite-difference scheme in configuration space.}. The discrete Fourier transform in $\{x,y\}$, 

\begin{equation}
	g_{\s} (x,y,z, v_{\parallel}, \mu_{\s},t) = \sum_{k_x, k_y} \hat{g}_{\boldsymbol{k},\s} (z,v_{\parallel}, \mu_\s,t) e^{(ik_x x+ i k_y y)} \; ,
	\label{eq:fourier_transform}
\end{equation}

\noindent
is justified provided the condition $k_x \sim k_y \gg 1/L$ is satisfied. This translates to a requirement that the turbulent fluctuations at the edges of our domain are decorrelated, such that they may be considered statistically identical, and periodic boundary conditions can be enforced in $(x,y)$ (\citealt{Beer95}).

\subsection{System Equations} \label{sec:system_equations}
\noindent
We next transform our evolution equations \eqref{eq:full_EM_gyrokinetic_eqn}, and \eqref{eq:quasineutrality_unnormalised}-\eqref{eq:perp_ampere_law_unnormalised}. The definition of $\chi$ is used to find the Fourier transform of the quantity $ \left\langle \chi \right\rangle_{\gyror}  = \langle \phi \rangle_{\gyror} - v_{\parallel}  \langle A_{\parallel} \rangle_{\gyror}/c - \langle \boldsymbol{v}_{\perp} \cdot \vec{A}_{\perp} \rangle_{\gyror}/c  $:
\begin{align}
	\mathcal{F}_{\vec{k}} \left[ \left\langle \chi \right\rangle_{\gyror} \right] \doteq \gyrochi  = \left[ \besselk \pk - \frac{v_{\parallel}}{c} \besselk \ak  + 2 \frac{\mu_{\s}}{Z_{\s} e} \frac{J_{1,\boldsymbol{k},\s} }{a_{\boldsymbol{k},\s }} \bk \right] ,
	\label{eq:gyro_chi_unnormalised}
\end{align}
\noindent
where we define $	\mathcal{F}_{\vec{k}} \left[ \left\langle \chi \right\rangle_{\gyror} \right] \doteq \gyrochi$ to be the Fourier components of $ \left\langle \chi \right\rangle_{\gyror}$. The variables $J_{n, \boldsymbol{k}, \s}$ above are the n\textsuperscript{th}-order Bessel functions of the first kind for species $\s$. The Bessel functions arise naturally as a result of the gyroaverages that appear in \eqref{eq:full_EM_gyrokinetic_eqn}, and have argument $a_{\boldsymbol{k}, \s} = k_{\perp} v_{\perp} / \Omega_{\s}$. Utilising this, we can write each of our functional operators in terms of $\{ \gks, \pk, \ak, \bk \}$. The resulting gyrokinetic equation is
\begin{align}
    \hat{G}_{\boldsymbol{k},\s} = & \; \pdv{\gks}{t} + v_{\parallel} \hat{\boldsymbol{b}} \cdot \grad z \left[\pdv{\gks} {z} + \frac{Z_{\s} e}{T_{\s}}  \pdv{\gyrochi}{z} \maxwellian \right] \nonumber \\
	& + i \omega_{*,\boldsymbol{k},\s} \maxwellian \gyrochi + i \omega_{d,\boldsymbol{k},\s} \left[ \gks +  \frac{Z_{\s} e}{T_{\s}}  \gyrochi \maxwellian \right] \nonumber \\
	& -  \frac{\mu_{\s}}{m_{\s}} (\hat{\boldsymbol{b}} \cdot \grad B_0) \pdv{\gks}{v_{\parallel}}  + \frac{Z_{\s} e}{T_{\s}} \frac{\mu_{\s}}{m_{\s} c } (\hat{\boldsymbol{b}} \cdot \grad B_0) \maxwellian \besselk \ak - \hat{C}_{\boldsymbol{k},\s} [\gks]
	\nonumber \\ 
	= & \; 0.
	\label{eq:gyro_unnomalised_decomposed}
\end{align}
\noindent
The drift frequencies, $\omega_{d,\s}$ and $\omega_{*,\s}$, correspond to the magnetic drift frequencies resulting from the gradient and curvature of the magnetic field and the diamagnetic drift respectively, and are given by
\begin{align}
	\omega_{d, \boldsymbol{k}, \s} &= \frac{1}{\Omega_{\s}} (v_{\parallel}^2 \boldsymbol{v}_{\kappa} + \mu_{\s} \boldsymbol{v}_{\nabla B} ) \cdot ( k_x \grad x + k_y \grad y ) ,
	\label{eq:omega_d_def_unnormalised}\\
	\omega_{*,\boldsymbol{k}, \s} &= \frac{c k_y } {B_{0} } \frac{d y}{d \alpha} \frac{d \ln \maxwellian}{d \psi} ,
	\label{eq:omega_star_def_unnormalised}
\end{align}
\noindent
with $ \boldsymbol{v}_{\kappa} = \hat{\vec{b}} \cross ( \hat{\vec{b}} \cdot \grad \hat{\vec{b}} )$, and $ \boldsymbol{v}_{\nabla B} = \hat{\vec{b}} \cross \grad B $.
The distribution function and field quantities are time-dependent functions and will experience an initial transient period followed by exponential growth or decay. These equations admit normal mode solutions. Therefore we decompose $\hat{f}_{\boldsymbol{k}} = \sum_{j} \bar{f}_{\boldsymbol{k},j} e^{\gamma_{\boldsymbol{k},j} t} $\footnote{We remind the reader that here $\bf{k}$ denotes the Fourier mode from the spatial decomposition, and $j$ is the subscript denoting the temporal normal mode.} with $\gamma_{\boldsymbol{k},j} \in \mathbb{C}$ the complex frequencies, and $\bar{f}_{\boldsymbol{k},j}$ the amplitude for the $j$\textsuperscript{th} normal mode. Quasineutrality and Amp\`{e}re's law, along with the linearity of the gyrokinetic equation, ensure that all fluctuating quantities for an equivalent normal mode share the same complex frequency and will thus exhibit the same time-dependent behaviour during the period of exponential growth or decay. 
The initial transient period arises due to the competition between various normal modes that may have similar initial amplitudes. After a sufficiently long period of time (the exact length of which will depend on both the relative growth rates and starting amplitudes of each mode) the fastest growing mode will dominate, meaning we can approximate the time dependent behaviour after large times using a single temporal mode; $\hat{f}_{\boldsymbol{k}} \approx \bar{f}_{\boldsymbol{k},0} e^{\gamma_{\boldsymbol{k}, 0} \tilde{t}}$, with $\mathfrak{R}(\gamma_{\boldsymbol{k},0}) > \mathfrak{R}(\gamma_{\boldsymbol{k},j})$, $\forall j$, $j \neq 0$. Using this normal mode decomposition in the gyrokinetic equation and considering behaviour beyond the transient period gives
\begin{align}
    \hat{G}_{\boldsymbol{k},\s} = & \; \gamma_{\boldsymbol{k}, 0} \gksz+ v_{\parallel} \hat{\boldsymbol{b}} \cdot \grad z \left[\pdv{\gksz} {z} + \frac{Z_{\s} e}{T_{\s}}  \pdv{\chikz}{z} \maxwellian \right] \nonumber \\
	& + i \omega_{*,\boldsymbol{k},\s} \maxwellian \chikz + i \omega_{d,\boldsymbol{k},\s} \left[ \gksz +  \frac{Z_{\s} e}{T_{\s}}  \chikz \maxwellian \right] \nonumber \\
	& -  \mu_{\s} \hat{\boldsymbol{b}} \cdot \grad B_0 \pdv{\gksz}{v_{\parallel}}  + \frac{Z_{\s} e}{T_{\s}} \frac{\mu_{\s}}{m_{\s} c } (\hat{\boldsymbol{b}} \cdot \grad B_0) \maxwellian \besselk \akz - \hat{C}_{\boldsymbol{k},\s} [\gksz].
	\label{eq:steady_gyro_eqn_EM_unnormalised}
\end{align}
\noindent
The corresponding transformed, normalised field equations are $\hat{Q}_{\vec{k}} = \hat{M}_{\vec{k}} = \hat{N}_{\vec{k}} = 0$, where
\begin{align}
	\hat{Q}_{\boldsymbol{k}} = & 
	\sum_{\s} Z_{\s} e \left\{  \frac{2 \pi B_0}{m_{\s} } \int{\rmd^2 v \; \besselk \gksz } + \frac{Z_{\s} e n_{\s} }{T_{\s}} \left( \Gamma_{0, \boldsymbol{k}, \s}  - 1 \right) \; \pkz + \frac{n_{\s} }{B_0} \Gamma_{1, \boldsymbol{k}, \s} \bkz \right\} ,
	\label{eq:steady_quasi_EM_unnormalised}\\ 
	\hat{M}_{\boldsymbol{k}} = & \; - \frac{4 \pi}{k_{\perp}^2  c} \sum_{\s} Z_{\s} e \frac{2 \pi B_0}{m_{\s} }  \int \rmd^2 v \; v_{\parallel} \besselk \gksz
	\nonumber \\
	& + \left[1 + \frac{4 \pi}{ k_{\perp}^2  c^2} \sum_{\s} \frac{ (Z_{\s} e)^2 n_{\s}}{m_{\s}} \Gamma_{0,\boldsymbol{k}, \s} \right] \akz \; ,
	\label{eq:steady_par_amp_EM_unnormalised}\\ 
	\textnormal{and,} \; \; & \nonumber
	\\
	\hat{N}_{\boldsymbol{k}} = & \; 8 \pi \sum_{\s} \frac{2 \pi B_0}{m_{\s} } \int \rmd^2 v \frac{J_{1, \boldsymbol{k}, \s} }{a_{\boldsymbol{k}, \s}} \mu_{\s} \gksz + \left[ 4 \pi \sum_{\s} \frac{Z_{\s} e n_{\s}}{B_0} \Gamma_{1,\boldsymbol{k}, \s} \right] \pkz \nonumber \\
	& + \left[1+ 16 \pi \sum_{\s} \frac{ n_{\s} T_{\s}}{B_0^2} \Gamma_{2,\boldsymbol{k}, \s} \right] \bkz , 
	\label{eq:steady_perp_amp_EM_unnormalised}
\end{align}

\noindent 
with $ \int \rmd^2 v \doteq \int \rmd \mu \int \rmd v_{\parallel} $. In the above we have defined 
\begin{align}
	\Gamma_{0,\vec{k}, \s} & =  I_0 (\alpha_{\vec{k},\s} ) e^{- \alpha_{\vec{k},\s} }  \nonumber \\
	\Gamma_{1,\vec{k},\s} & = [I_0(\alpha_{\vec{k},\s} ) - I_1(\alpha_{\vec{k},\s} ) ] e^{- \alpha_{\vec{k},\s} } \nonumber \\
	\Gamma_{2,\vec{k},\s} & = I_1 (\alpha_{\vec{k},\s}) e^{- \alpha_{\vec{k},\s} } \nonumber 
\end{align}
where $I_0$ and $I_1$ are modified Bessel functions of the first kind, and $\alpha_{\vec{k}, \s} = k_{\perp}^2 \rho_{\s}^2 /2 $.

\section{Adjoint Method for Gyrokinetics}\label{sec:Adjoint_method_GK}
\noindent
The adjoint-based optimisation method is a tool that, at its heart, efficiently calculates derivatives of a desired quantity with respect to a potentially large number of parameters. 
The associated computational cost depends only on the expense of solving both the objective function and adjoint system of equations, and is essentially independent of the dimension of the parameter space. Adjoint methods have already been successfully applied to certain other aspects of MCF devices, such as optimising coil configurations for stellarator geometries (see, e.g., \citealt{Paul18}; \citealt{geraldini_landreman_paul_2021}; \citealt{Nies22}). The novelty here is to apply the adjoint method to geometric optimisation and plasma microstability, which includes the complexity of the full linearised gyrokinetic system.  

We now turn to the gyrokinetic equation and the objective of minimising the dominant linear growth rate, $\gamma_{\boldsymbol{k}, 0}$, with respect to a set of currently unspecified parameters $\{ p_i \}$, which we take to be the components of vector $\bp$. This section outlines how one can take advantage of the adjoint method to efficiently obtain the gradient $\grad_{\vec{p}} \gamma_{\boldsymbol{k}, 0}$. Although a finite difference scheme could be used to obtain such a gradient this becomes computationally expensive when the dimension of $\vec{p}$, $\mathcal{N}_{\vec{p}}$, is large: for each gradient a finite difference scheme demands we solve the system equations $\mathcal{N}_{\vec{p}} + 1 $ times. In contrast, the adjoint method allows us to solve the system equations only once, and in exchange one must solve the set of adjoint equations, for which the cost is computationally equivalent to the original system equations. 

We start by considering the general case of subsonic, linear, electromagnetic, $\delta \! f$-gyrokinetics including collisions, with equations \eqref{eq:steady_gyro_eqn_EM_unnormalised}-\eqref{eq:steady_perp_amp_EM_unnormalised} as the functional operators defining our system in the long time limit.

\subsection{General Formalism} \label{sec:EM_adjoint_overview}
Since our equations are limited to the linear regime, there is no coupling of different Fourier modes and it is thus possible to consider each perpendicular wavenumber individually. Given that we are also considering the post-transient limit with only one dominant growth rate, it will henceforth be assumed that only a single perpendicular wavenumber is being considered, and we will thus drop the $\boldsymbol{k}$ subscript, along with the subscript that denotes the dominant mode for the distribution function and field quantities. For notational brevity we shall also drop the over-bars that appear on the distribution function and fields that denoted normal mode decomposition. Hence everywhere $\gun$, $\pun$, $\aun$, and $\bun$ appear it shall be assumed that they contain suppressed spatial and temporal Fourier subscripts. 

Consider now a set of variables that influences our system and which exists within a parameter space spanned by all possible $\bp$. We take the set $\{ p_i\}$, $i \in [1, \mathcal{N}_p ]$, to be linearly independent, with no time variation, and explore how variations within this space affect the linear growth rate. 
At present we need not specify which variables are denoted by $\bp$, and thus derive a general set of adjoint equations for the gyrokinetic-Maxwell system above, \eqref{eq:steady_gyro_eqn_EM_unnormalised}-\eqref{eq:steady_perp_amp_EM_unnormalised}. 

Our objective function, $\hat{G}_{\s}$, is a linear function of $\{\gun, \pun, \aun, \bun \}$, which are coupled to the fields through the field equations, \eqref{eq:steady_quasi_EM_unnormalised}-\eqref{eq:steady_perp_amp_EM_unnormalised}. When taking derivatives of $G_{\s}$ we invariably end up with derivatives acting on all four of these variables. This is undesirable because it requires we calculate the gradients of $\hat{g}_{\s}$ and the fields, which in turn requires us to solve the gyrokinetic system $\mathcal{N}_p + 1$ times, as discussed above. In order to eliminate these four derivatives we use something akin to the method of Lagrange multipliers and introduce four adjoint variables which multiply the corresponding constraint equations. The optimisation Lagrangian is thus
\begin{equation}
    \mathcal{L} \coloneqq \inprodzv{\hat{G}_{\s}}{\lamspec} + \inprodz{\hat{Q}}{\xi}+ \inprodz{\hat{M}}{\zeta }+ \inprodz{\hat{N}}{\sigma} ,
	\label{eq:EM_optimisation_lag}
\end{equation}
\noindent
with the angle brackets representing inner products defined through: 
\begin{align}
	\inprodz{a}{b}  & = \int{ \frac{\rmd z }{B_0 \; \bdotgradzun}  \; a \;  b^* } , \quad
	\inprodv{a}{b} = \sum_{\s} \frac{2 \pi B_0}{m_{\s} } \int \rmd^2 v \; a \; b^* , \nonumber \\
	\inprodzv{a}{b} & = \sum_{\s} \int \frac{\rmd z}{B_0 \; \bdotgradzun } \; \frac{2\pi B_0}{m_{\s} } \int \rmd^2 v \; a \; b^* .
	\label{eq:inner_producs}
\end{align}

\noindent
We have introduced the set of adjoint variables  $\lamspec$, $\xi$, $\zeta$, and $\sigma$, whose forms are to be determined\footnote{Note that the adjoint variables here are defined in Fourier space, so $\lamspec$, $\xi$, $\zeta$, and $\sigma$ also contain suppressed $\boldsymbol{k}$ subscripts, and thus we are calculating each for a specific $\boldsymbol{k}$. The derivatives $\gradp$ indicate how these Fourier components respond to external changes of the parameters $\bp$ in our system.}. We identify $\lamspec$ as the adjoint variable to the distribution function, $\gun$, whereas $\xi$, $\zeta$, and $\sigma$ are adjoint to the field variables (referred to henceforth as adjoint fields). The quantities $\hat{G}_{\s}$, $\hat{Q}$, $\hat{M}$, and $\hat{N}$ are our objective functions, and we remark that since $\hat{G}_{\s}= \hat{Q} = \hat{M} = \hat{N} = 0$ for a consistent set of $\{ \bp_0, \gun(\bp_0), \pun(\bp_0), \aun (\bp_0), \bun (\bp_0)\}$ we have $\left. \mathcal{L}\right|_{\bp_0} = 0$ for all choices of adjoint variables. 
For later convenience we decompose the functional operators into their components that act on $\gun$, $\pun$, $\aun$, and $\bun$ separately:
\begin{align}
	\hat{G}_{\s} [\boldsymbol{p}; \gun, \pun, \aun, \bun] &
	\!\begin{aligned}[t]
		= \; & \hat{G}_{g, \s} [\bp ; \gun] + \hat{G}_{\phi, \s} [\bp ; \pun] + \hat{G}_{A_{\parallel}, \s} [\bp ; \aun] + \hat{G}_{B_{\parallel}, \s} [\bp ;\bun] \\
		& - \hat{C}_{\s}[\bp;\gun]
	\end{aligned}
	\label{eq:GK_decomposition_EM}\\
	\hat{Q} [\bp ; \gun, \pun, \bun] & = \inprodv{\hat{Q}_{g, \s} [\bp ; \gun]}{\mathbb{I}} +  \hat{Q}_{\phi} [\bp ; \pun ] + \hat{Q}_{B_{\parallel}} [\bp ; \bun]
	\label{eq:quasi_decomposition_EM}\\
	\hat{M} [\bp ; \gun, \aun] & = \inprodv{\hat{M}_{g,\s} [\bp ; \gun] }{\mathbb{I} } + \hat{M}_{A_{\parallel}} [\bp ; \aun ] 
	\label{eq:par_amp_decomposition_EM}\\
	\hat{N} [\bp ; \gun, \pun, \bun ] & = \inprodv{\hat{N}_{g,\s} [\bp ; \gun ]}{\mathbb{I}}  + \hat{N}_{\phi} [\bp ; \pun ] + \hat{N}_{B_{\parallel}} [\bp; \bun] ,
	\label{eq:perp_amp_decomposition_EM}
\end{align}
\noindent
with $\mathbb{I}$ the identity matrix. Explicit expressions for these operators are given in \apref{sec:decomp_operators}. 

We next consider taking the gradient derivative of \eqref{eq:EM_optimisation_lag} with respect to the variables in our parameter space, $\bp$. We now isolate all terms multiplying derivatives of $\{\gun, \pun, \aun, \bun \}$ so that each of their coefficients can be set to zero. We are at liberty to do this because of the freedom that exists in choosing the adjoint variables introduced in \ref{eq:EM_optimisation_lag}. The gradient derivative of \eqref{eq:EM_optimisation_lag} is expanded using equations \eqref{eq:GK_decomposition_EM}-\eqref{eq:perp_amp_decomposition_EM}:
\begin{align}
	\gradp \mathcal{L} = & \; \partial_{\bp} \mathcal{L} + \inprodzv{\gradp \gun } {  \hat{G}^{\dag}_{g,\s} [\bp; \lamspec] - \hat{C}^{\dag}_{\s} [\bp ; \lamspec] + \hat{Q}^{\dag}_{g,\s} [\bp; \xi] +  \hat{M}^{\dag}_{g,\s} [\bp; \zeta]  +  \hat{N}^{\dag}_{g,\s} [\bp; \sigma]} 
	\nonumber \\
	& + \inprodz { \gradp \pun } { \hat{G}^{\dag}_{\phi,\s} [\bp ; \lamspec] + \hat{Q}^{\dag}_{\phi} [\bp ; \xi] + \hat{N}^{\dag}_{\phi} [\bp ; \sigma] }
	+ \inprodz { \gradp \aun }  {\hat{G}^{\dag}_{A_{\parallel},\s} [\bp ; \lamspec] + \hat{M}^{\dag}_{A_{\parallel}} [\bp ; \zeta]}
	\nonumber \\ 
	& + \inprodz {  \gradp \bun } { \hat{G}^{\dag}_{B_{\parallel},\s} [\bp ; \lamspec] +  \hat{Q}^{\dag}_{B_{\parallel}} [\bp ; \xi] + \hat{N}^{\dag}_{B_{\parallel}} [\bp ; \sigma] } + \mathcal{B} .
	\label{eq:derivative_of_optimisation_Lagrangian}
\end{align}

\noindent
Here the partial derivative, $\partial_{\bp}$, is taken at fixed $\gun$, $\pun$, $\aun$, and $\bun$. Note that it acts on the inner product itself in addition to the terms within it to account for the $\bp$-dependence of the Jacobians\footnote{It is noteworthy to point out that although the Jacobians present in the integrals (as well as the Lagrange multipliers themselves) are $\bp$-dependent, we anticipate that when the distribution function and fields are evaluated at $\bp_0$ their derivatives provide zero contribution as they multiply the functional operators, which are identically zero at $\bp_0$. Hence, we would be justified in pulling the partial derivative through these inner products, and the final derivative of the growth rate obtained should be the same.} and functional operators. The term \enquote*{$\mathcal{B}$} accounts for boundary terms that arise when we integrate by parts to invert operators such as $\inprodz {\gradp \partial_z \gun }{\lamspec}$ onto the adjoint variables, $\inprodz {\gradp \gun}{\partial_z \lamspec}$, and is given by
\begin{align}
	\mathcal{B} = & \sum_{\s}  \frac{2 \pi B_0}{m_{\s} } \int \rmd^2 v \; v_{\parallel} \; \lamspec^* \left[ \gradp \gun + \frac{Z_{\s} e}{T_{\s} } \bessel \maxwellian \gradp \pun 
	\right. \nonumber \\
	& \left. \left.- 2 \frac{Z_{\s} e}{T_{\s} } v_{\parallel} \bessel \maxwellian \gradp \aun + 4 \mu_{\s} \frac{J_{g,\s}}{a_{\s}} \maxwellian \gradp \bun  \right] \right|_{z = -\infty}^{z = \infty} \nonumber 
	\\
	& \left. - \sum_{\s} \frac{2\pi B_0}{m_{\s} } \int \rmd z \int \rmd \mu \; \mu_{\s} \pdv{B_0}{z} \lamspec^* \gradp \gun \right|_{v_{\parallel} = - \infty}^{v_{\parallel} = \infty} .
\end{align}

\noindent
We can conveniently set these terms to zero by applying appropriate restrictions on our adjoint variables. These terms consequently define the boundary conditions we apply to our adjoint variables. The incoming boundary conditions along the magnetic field on $\gun$ are taken to be $\gun(z\rightarrow -\infty, v_{\parallel} > 0, \mu_{\s}), \; \gun(z \rightarrow \infty, v_{\parallel} <0, \mu_{\s}) \rightarrow 0$, independently of $\bp$, such that $d_{\bp} \gun = 0$ at these limits. We impose the boundary condition on $\lamspec^*$ to be $ \lamspec^* (z \rightarrow -\infty, v_{\parallel} <0, \mu_{\s}), \; \lamspec^* (z \rightarrow \infty,v_{\parallel}>0, \mu_{\s}) \rightarrow 0$ in order to eliminate the boundary term arising from the $z$ integration by parts, and hence remove the need to calculate $\gradp \{\gun,\pun,\aun,\bun \}$ at the boundaries in $z$. The boundary term arising from integration by parts in $v_{\parallel}$ is automatically satisfied as it is assumed that $\gun(z, v_{\parallel} \rightarrow \pm \infty, \mu_{\s}) = 0$, $\forall \{z, \mu_{\s} \}$ independently of $\bp$. However it is convenient to impose that $\lamspec^* (z, v_{\parallel} \rightarrow \pm \infty, \mu_{\s}) = 0 $ such that $\lamspec^*$ and $\gun$ satisfy similar boundary conditions, whilst also ensuring $\lamspec$ is sensibly defined and normalisable\footnote{An additional consequence is that this choice simplifies the implementation into an existing gyrokinetic code.}. 
The substitution $\tildlamconj = \lamspec^*(z,-v_\parallel,\mu_{\s})$ is made such that the adjoint equations more closely resemble those in the original gyrokinetic system. This redefines the $z$-boundary condition on the adjoint variable $\tildlamconj (z \rightarrow - \infty, v_{\parallel} > 0, \mu_{\s}), \; \tildlamconj (z \rightarrow \infty, v_{\parallel} < 0, \mu_{\s}) \rightarrow 0$, which now mirrors those satisfied by $\gun$.

Setting the remaining coefficients of $\gradp \gun$,  $\gradp \pun$, $\gradp \aun$, and $\gradp \bun$ in \eqref{eq:derivative_of_optimisation_Lagrangian} equal to zero yields the constraint equations for the adjoint variables
\begin{align}
	\hat{G}^{\dag}_{g,\s} [ \bp ; \tildlam] + \hat{Q}^{\dag}_{g,\s} [\bp ; \xi] + \hat{M}^{\dag}_{g,\s} [\bp; \zeta ] + \hat{N}^{\dag}_{g,\s} [\bp ; \sigma] - \hat{C}^{\dag}_{\s} [\bp ; \tildlam] & = 0 \; ,
	\label{eq:lam_symbolic_eq}
	\\
	\langle \hat{G}^{\dag}_{\phi,\s} [\bp ; \tildlam ] \rangle_{v, {\s}} + \hat{Q}^{\dag}_{\phi} [ \bp ; \xi ] + \hat{N}^{\dag}_{\phi} [\bp ; \sigma] & = 0 \; , 
	\label{eq:xi_symbolic_eq}
	\\
	\langle \hat{G}^{\dag}_{A_{\parallel},\s} [ \bp ; \tildlam] \rangle_{v, {\s}} + \hat{M}^{\dag}_{A_{\parallel}} [\bp ; \zeta] & = 0 \; ,
	\label{eq:zeta_symbolic_eq}\\
	\langle \hat{G}^{\dag}_{B_{\parallel},\s} [ \bp ; \tildlam] \rangle_{v, {\s}} + \hat{Q}^{\dag}_{B_{\parallel}} [\bp ; \xi] + \hat{N}^{\dag}_{B_{\parallel}} [ \bp ; \sigma] & = 0 \; .
	\label{eq:sigma_symbolic_eq}
\end{align}

\noindent
The expressions for these adjoint operators are stated in \apref{sec:adjoint_operators}. The $z$-derivatives which appear in the $\hat{G}_{\s}$ operators, under the velocity integrals, in equations \eqref{eq:xi_symbolic_eq}-\eqref{eq:sigma_symbolic_eq} pose a potential difficulty; to calculate the adjoint fields, information is required for $\lamspec^*$ at all $z$, for both positive and negative velocities. Given the boundary conditions on $\lamspec$ and the propagation of information by advection this information is not readily available.  In order to circumvent this problem, moments of \eqref{eq:lam_symbolic_eq} are taken to simplify the equations. A summary of this calculation can be found in \apref{sec:Simplifying_adjoint_eqns}, and the result after algebraic manipulation is:
\begin{align}
	& \!\begin{aligned}[b]
	\gamma^* \tildlam & + v_{\parallel} \bdotgradzun \pdv{\tildlam}{z}
	- \frac{\mu_{\s}}{m_{\s}} \bdotgradzun \pdv{B_0}{z} \pdv{\tildlam}{v_{\parallel}} - i \omega_{d, \s} \tildlam + Z_{\s} e \bessel \xi \\
	& - \frac{4 \pi}{k_{\perp}^2} Z_{\s} e \bessel \frac{v_{\parallel}}{c}  \zeta + 8 \pi \frac{J_{1,\s}}{a_{\s}} \mu_{\s} \sigma - \hat{C}_{\s}[\tildlam] = 0 \; ,
	\end{aligned}
	\label{eq:final_lam_eqn_EM}
	\\
	&\xi  + \frac{1}{\eta} \sum_{\s} \frac{2 \pi B_0}{m_{\s} } \int \rmd^2 v \left[i \omega_{*,\s} + \frac{Z_{\s}e}{T_{\s}} \gamma^* \right] \bessel \maxwellian \tildlam =0 \; ,
	\label{eq:final_xi_eqn_EM}
	\\
	& \zeta - \frac{1}{k_{\perp}^2 } \sum_{\s} \frac{2 \pi B_0}{m_{\s} } \int \rmd^2 v \; \frac{v_{\parallel}}{c}  \left[ i \omega_{*,\s} + \frac{Z_{\s}e}{T_{\s}} \gamma^* \right] \bessel \maxwellian \tildlam = 0 \; ,
	\label{eq:final_zeta_eqn_EM}
	\\
	&\sigma -  \sum_{\s} \frac{2 \pi B_0}{m_{\s} }  \int \rmd^2 v \; \left( 2 \frac{\mu_{\s} }{Z_{\s} e}\frac{J_{1,\s}}{a_{\s}} \right) \left[ i \omega_{*,\s} + \frac{Z_{\s} e}{T_{\s} } \gamma^* \right] \maxwellian \tildlam = 0 \; ,
	\label{eq:final_sigma_eqn_EM}
\end{align}

\noindent 
with $ \eta = \sum_{\s} (Z_{\s} e)^2 n_{\s}/ T_{\s} $. Noting that we can also rewrite $\hat{G}_{\s} [\bp; \gun, \pun, \aun, \bun]= \gamma \gun + \hat{L}_{\s} [\bp; \gun, \pun, \aun, \bun]$, and using $\left. \grad_{\bp} \mathcal{L} \right|_{\bp_0} = 0$, we can rearrange equation \eqref{eq:derivative_of_optimisation_Lagrangian} to obtain
\begin{align}
	\gradp \gamma \; \langle \gun, \lamspec \rangle_{z,v, {\s}} = - \left. \left[ \langle \partial_{\bp} \hat{L}_{\s} , \lamspec \rangle_{z,v, {\s}} + \langle \partial_{\bp} \hat{Q}, \xi \rangle_{z} + \langle \partial_{\bp} \hat{M} , \zeta \rangle_{z} + \langle \partial_{\bp} \hat{N}, \sigma \rangle_{z} \right]  \right|_{\bp_0} \; ,
	\label{eq:final_gam_eqn_EM}
\end{align}

\noindent 
where the partial derivatives have been pulled inside the inner products as the contribution arising from the Jacobian derivatives is zero by virtue of $\hat{G}_{\s}(\bp_0) = \hat{Q}(\bp_0) = \hat{M}(\bp_0) = \hat{N}(\bp_0) = 0$. We can then solve \eqref{eq:final_gam_eqn_EM} using the closure equations provided by \eqref{eq:final_lam_eqn_EM}-\eqref{eq:final_sigma_eqn_EM}.
\subsection{Electrostatic, Collisionless Limit} \label{sec:electrostatic_adjoint}
\noindent
We consider now the electrostatic, collisionless limit, as we shall perform numerical tests in this regime. In the limit of small plasma $\beta$, meaning the ratio of plasma to magnetic pressure tends to zero, the magnetic field perturbations tend to zero. Hence, to extract the electrostatic, collisionless limit from these equations we set $\aun = \bun = 0$, and $C_{\s, \s'} =0$. The linear, collisionless, electrostatic gyrokinetic equation is:
\begin{equation}
	\begin{split}
		\hat{G}_{\s} [\bp; \gun, \pun] = &  \; \gamma \gun + v_{\parallel} \bdotgradzun \left[\pdv{\gun}{z} + \frac{Z_{\s} e}{T_{\s} } \pdv{\bessel \pun}{z} \maxwellian \right] + i \omega_{*,\s} \bessel \pun \maxwellian \\ & - \frac{\mu_{\s}}{m_{\s} } \bdotgradzun \pdv{B_0}{z} \pdv{\gun}{v_{\parallel}} 
		+ i \omega_{d,\s} \left[ \gun + \frac{Z_{\s} e}{T_{\s} } \bessel \pun \maxwellian \right],
		\label{eq:steady_gyro_eqn_electrostatic}
	\end{split}
\end{equation}

\noindent
which is closed by the electrostatic limit of quasineutrality:
\begin{equation}
	\hat{Q} [\bp;\gun, \pun] = \sum_{\s}Z_{\s} e \left[  \frac{2\pi B_0}{m_{\s} } \int d^2 v \; \bessel \gun + \frac{Z_{\s} n_{\s}}{T_{\s}} (\Gamma_{0,\s} -1) \pun \right] ,
	\label{eq:quasineutrality_eqn_electrostatic}
\end{equation}

\noindent
with $\hat{G}_{\s}$ and $\hat{Q}$ identically zero. In the above we are once again considering the long time behaviour of a single wavenumber, and have suppressed the associated subscripts. 

As in \secref{sec:EM_adjoint_overview} we decompose the functional operators $\hat{G}_{\s}[\bp; \gun, \pun ]$, and $\hat{Q}[\bp; \gun, \pun]$ into components that act on $\gun$ and $\pun$ separately, with all other operators in \eqref{eq:GK_decomposition_EM}-\eqref{eq:perp_amp_decomposition_EM} set to zero. The derivation in \secref{sec:EM_adjoint_overview} is unchanged, with the exception that some terms may now be omitted. The resulting derivative of the growth rate in the electrostatic, collisionless regime is:
\begin{align}
	\gradp \gamma \; \langle \gun, \lamspec \rangle_{z,v, {\s}} = & - \left. \left[  \langle \partial_{\bp} \hat{L}_{\s}, \lamspec \rangle_{z,v, {\s}}  + \langle \partial_{\bp} \hat{Q}, \xi \rangle_{z} \right] \right|_{\bp_0},
	\label{eq:electrostatic_derivative}
\end{align}

\noindent
where $\hat{L}_{\s}= \hat{G}_{\s} - \gamma g_{\s}$ is given by equation \eqref{eq:steady_gyro_eqn_electrostatic}, and the adjoint equations are

\begin{align}
	\gamma^* \tildlam + v_{\parallel} \bdotgradzun \pdv{\tildlam}{z}
	- \frac{\mu_{\s}}{m_{s}} \bdotgradzun \pdv{B_0}{z}  \pdv{\tildlam}{v_{\parallel}} - i \omega_{d, \s} \tildlam
	+ Z_{\s} e \bessel \xi & = 0 \; ,
	\label{eq:final_lam_eqn_electrostatic}
	\\
	\xi  + \frac{1}{\eta} \sum_{\s} \frac{2 \pi B_0} {m_{\s} } \int \rmd^2 v \left[i \omega_{*,\s} + \frac{Z_{\s} e}{T_{\s} } \gamma^* \right] \bessel \maxwellian \tildlam & = 0 \; ,
	\label{eq:final_xi_eqn_electrostatic}
\end{align}

\noindent
with $\tildlam (v_{\parallel} ) = \lamspec(-v_{\parallel} )$, and $ \eta = \sum_{\s} (Z_{\s} e)^2 n_{\s}/ T_{\s} $ as before.

\section{Normalisations and Magnetic Geometry} \label{sec:implementation_into_code}
\noindent
To evaluate \eqref{eq:electrostatic_derivative}, we need to solve for the set of variables $\{\gamma, \gun, \pun, \tildlam, \xi \}$, evaluated at the unperturbed geometric values, $\bp_0$. We do this by implementing and combining the adjoint system within the local $\delta \! f$-gyrokinetic code \texttt{stella} \citep{Barnes19}. In this section we write the gyrokinetic equations in normalised coordinates along with the corresponding normalised equations for the adjoint variables. We then introduce a specific choice for $\bp$, the Miller parametrisation used to specify local magnetic equilibria in tokamaks \citep{Miller98}, and detail how the adjoint method can be applied in this case.

\subsection{Normalisation} \label{sec:Normalisation}
\subsubsection{Gyrokinetic Normalisations} \label{sec:gk_normalisations}
\noindent
Here we normalise the full electromagnetic gyrokinetic system and give the reduced electrostatic limit at the end of the section. We choose our reference quantities to be the same as in \texttt{stella} to aid in numerical implementation, and denote these with a subscript \enquote*{r}. A reference length scale that characterises the parallel lengths in the simulation is introduced as $L=a$; for the Miller formalism \texttt{stella} takes $a$ to be the half diameter of the plasma volume (minor radius), at the height of the magnetic axis. The perpendicular reference length scale is taken as a reference gyroradius $\rho_r \doteq v_{th,r}/\Omega_r$, with $v_{th,r} = \sqrt{2 T_r/m_r}$ the reference velocity, and $\Omega_r = e B_r/m_{r}c$, with $T_r$, $n_r$, $B_r$, and $m_r$ being user specified quantities. The normalised variables are denoted with a tilde, and are provided in Table~\ref{tab:normalised_quantities}. 

Multiplying the gyrokinetic equation, rewritten in normalised coordinates, by the factor $(a^2/\rho_r v_{th,r}) \text{exp}(-v^2/v_{th,\s}^2)/F_{0,\s}$ we obtain the normalised low-flow, electromagnetic gyrokinetic equation taken in the long time limit
\begin{align}
    \hat{G}_{\boldsymbol{k},\s} = & \; \tilde{\gamma} \gk + \gradpar \left[\pdv{\gk}{\tilde{z}} + \zt \pdv{\left\langle \tilde{\chi} \right\rangle_{\vec{k}, \s}}{\tilde{z}} \maxwexp \right] \nonumber \\
	& + i \tilde{\omega}_{*,\boldsymbol{k},\s} \maxwexp \left\langle \tilde{\chi} \right\rangle_{\vec{k}, \s} + i \tilde{\omega}_{d,\boldsymbol{k},\s} \left[ \gk + \zt \left\langle \tilde{\chi} \right\rangle_{\vec{k}, \s} \; \maxwexp \right] \nonumber \\
	& - \vthnorm \tilde{\mu}_{\s} \hat{\boldsymbol{b}} \cdot \tilde{\grad} \tilde{B}_0 \pdv{\gk}{\tilde{v}_{\parallel}}  + 2 \frac{Z_{\s}}{\tilde{m}_{\s}} \tilde{\mu}_{\s} \hat{\boldsymbol{b}} \cdot \tilde{B}_0 \maxwexp \besselk \tilde{A}_{\parallel, \boldsymbol{k}} - \hat{C}_{\boldsymbol{k},\s} [\gk], 
	\label{eq:steady_gyro_eqn_EM_stella}
\end{align}

\begin{table}
	\begin{center} \setlength\tabcolsep{12.0pt} \caption{List of normalised parameters and variables} 	\label{tab:normalised_quantities} 
		\begin{threeparttable}\renewcommand{\arraystretch}{1.0}\footnotesize 
			\begin{tabular}{l c} \toprule\midrule
			\multicolumn{2}{l}{\emph{Normalised Parameters}} \\
			\midrule
			\emph{Parameter} & \emph{Normalisation}
			\\ \midrule 
			$\tilde{t}$ & $ t \: a/v_{th,r}$ \\
			$\tilde{T}_{\s}$ & $T_{\s}/T_{r}$ \\
			$\tilde{B} $ & $  B/ B_r$\\
			$\tilde{m}_{\s} $ & $  m_{\s} / m_r$\\
			$\tilde{n}_{\s} $ & $  n_{\s}/ n_{r}$ \\ 
			$\tilde{\grad} $ & $ a \grad$ \\
			$\tilde{v}_{\parallel} $ & $v_{\parallel}/v_{th,\s}$ \\ 
			$\tilde{\mu}_{\s} $ & $ \mu_{\s} B_r / m_{\s} v_{th,\s}^2 $ \\
			$\vthnorm $ & $  v_{th,\s}/v_{th,r}$ \\ 
			$\tilde{g}_{\boldsymbol{k}, \s}$ & $g_{\boldsymbol{k}, \s} (\maxwexp/\maxwellian) (a/\rho_r)$ \\
			$\tilde{\phi}_{\boldsymbol{k}} $ & $ \phi_{\boldsymbol{k}} (e/T_r) (a/\rho_r) $\\
			$\tilde{A}_{\parallel,\boldsymbol{k}} $ & $(a/ B_r \rho_r^2) A_{\parallel, \boldsymbol{k}}$ \\
			$\tilde{B}_{\parallel, \boldsymbol{k}} $ & $ (a/B_r \rho_r) \delta B_{\parallel, \boldsymbol{k}} $ \\
			$\tilde{a}_{\boldsymbol{k}, \s}$ & $ \tilde{k}_{\perp} \tilde{v}_{\perp} / \tilde{\Omega}_{\s}$ \\
			\midrule
			\multicolumn{2}{l}{\emph{Normalised Variables}} \\ \midrule
			$\left\langle \tilde{\chi} \right\rangle_{\vec{k}, \s} $ & \multicolumn{1}{l}{$ =
			\besselk \tilde{\phi}_{\boldsymbol{k}} - 2 \vthnorm \tilde{v}_{\parallel} \besselk \tilde{A}_{\parallel, \boldsymbol{k}} + 4 \tilde{\mu}_{\s} (\tilde{T}_{\s}/Z_{\s}) (J_{1,\boldsymbol{k},\s} / \tilde{a}_{\boldsymbol{k},\s }) \delta \tilde{B}_{\parallel, \boldsymbol{k}}$} \\
			$\tilde{\omega}_{d, \boldsymbol{k}, \s} $& \multicolumn{1}{l}{$= (\tilde{T}_{\s} \rho_r /Z_\s \tilde{B}) (\tilde{v}_{\parallel}^2 \boldsymbol{v}_{\kappa} + \tilde{\mu}_{\s} \boldsymbol{v}_{\nabla B} ) \cdot ( k_x \tilde{\grad} x + k_y \tilde{\grad} y ) $} \\
			$\tilde{\omega}_{*,\boldsymbol{k}, \s} $& \multicolumn{1}{l}{$ = (k_y \rho_r/2) a B_r (\rmd y / \rmd \alpha) (\rmd \ln \maxwellian / \rmd \psi)$ } \\
			\midrule \bottomrule
			\end{tabular} 
		\end{threeparttable} 
	\end{center} 
\end{table}


\noindent
The corresponding transformed, normalised field equations are given by 
\begin{align}
    \hat{Q}_{\boldsymbol{k}} = & 
    \sum_{\s} Z_{\s} \tilde{n}_{\s} \left\{  \frac{2\tilde{B}_0}{\sqrt{\pi}} \int{\rmd^2 \tilde{v} \; \besselk \gck } + \frac{Z_{\s}}{T_{\s}} \left( \Gamma_{0, \boldsymbol{k}, \s}  - 1 \right) \; \tilde{\phi}_{\boldsymbol{k},0} + \frac{1}{\tilde{B}_0} \Gamma_{1, \boldsymbol{k}, \s} \delta \tilde{B}_{\parallel, \boldsymbol{k}, 0} \right\} ,
	\label{eq:steady_quasi_EM_stella}\\ 
	\hat{M}_{\boldsymbol{k}} = & \; - \frac{\beta_r}{\left(k_{\perp}\rho_r\right)^2} \sum_{\s} Z_{\s} \tilde{n}_{\s} \vthnorm \velspace \int \rmd^2 \tilde{v} \; \tilde{v}_{\parallel} \besselk \gck
	\nonumber \\
	& + \left[1+ \frac{\beta_r}{ \left(k_{\perp}\rho_r\right)^2} \sum_{\s} \frac{Z_{\s} \tilde{n}_{\s}}{\tilde{m}_{\s}} \Gamma_{0,\boldsymbol{k}, \s} \right] A_{\parallel, \boldsymbol{k},0 }  \; ,
	\label{eq:steady_par_amp_EM_stella}\\ 
	\hat{N}_{\boldsymbol{k}} = & \; 2 \beta_r \sum_{\s} \tilde{n}_{\s} \tilde{T}_{\s}  \; \frac{ 2 \tilde{B}_0}{\sqrt{\pi}} \int \rmd^2 \tilde{v} \tilde{\mu}_{\s} \frac{\besselk}{\tilde{a}_{\boldsymbol{k}, \s}} \gck + \left[\frac{\beta_r}{2\tilde{B}_0} \sum_{\s} Z_{\s}\tilde{n}_{\s} \Gamma_{1,\boldsymbol{k}, \s} \right] \phick \nonumber \\
	& + \left[1+ \frac{\beta_r}{2\tilde{B}_0} \sum_{\s} Z_{\s}\tilde{n}_{\s} \tilde{T}_{\s} \Gamma_{2,\boldsymbol{k}, \s} \right] \delta \tilde{B}_{\parallel, \boldsymbol{k},0} ,
	\label{eq:steady_perp_amp_EM_stella}
\end{align}

\noindent
with the reference plasma beta, $\beta_r = 8\pi n_r T_r/B_r^2$. 
\subsubsection{Adjoint Normalisation} \label{sec:adjoint_normalisation}
\noindent
We choose the normalisation of the adjoint variables in such a way that our optimisation Lagrangian is dimensionless. In general this is achieved by enforcing that the dimension of the adjoint variables in \secref{sec:Adjoint_method_GK} satisfy:
\begin{align}
	[\lamspec] = [\hat{G}_{\s}]^{-1} \qquad [\xi] = [\hat{Q}]^{-1} \nonumber \\
	[\zeta] = [\hat{M}]^{-1} \qquad [\sigma ] = [\hat{N}]^{-1}
\end{align}
\noindent
with $[A]$ denoting the dimensionality of $A$. The normalised adjoint variables should then satisfy 
\begin{align}
	\tilde{\lambda}_{\s}  = \frac{\lamspec} {[\lamspec] } \qquad \xistella = \frac{\xi}{[\xi] } \nonumber \\
	\zetastella = \frac{\zeta}{[\zeta]} \qquad \sigstella = \frac{\sigma}{ [\sigma ] } .
\end{align} 
\noindent
Analysis of \eqref{eq:steady_gyro_eqn_EM_unnormalised}, and \eqref{eq:steady_quasi_EM_unnormalised}-\eqref{eq:steady_perp_amp_EM_unnormalised} gives the normalisations consistent with those chosen for the gyrokinetic variables
\begin{align}
	\tilde{\lambda}_{\s} = \lamspec \frac{\maxwellian} {\maxwexp } \frac{\rho_r v_{th,r} }{a^2} \qquad \xistella = \xi \frac{n_r e \rho_r}{ a }
	\nonumber \\
	\zetastella = \zeta \frac{ B_r \rho_r^2}{a} \qquad \sigstella = \sigma \frac{B_r \rho_r }{a} .
\end{align}
\noindent
We then multiply the adjoint equation \eqref{eq:final_lam_eqn_EM}, written in terms of normalised coordinates, by a factor of $ ( \rho_r/a ) \maxwellian/ \maxwexp  $ to obtain the electromagnetic, collisional adjoint equations in normalised units: 
\begin{align}
	& \!\begin{aligned}[b]
		\pdv{\steltildlam} {\tilde{t}} & + \tilde{\gamma}^* \steltildlam + \gradpar \pdv{\steltildlam}{\tilde{z}}
		- \dvpacoeff \pdv{\steltildlam}{\tilde{v}_{\parallel}} - i \tilde{\omega}_{d, \s} \steltildlam &  \\ 
		& + Z_{\s} \tilde{n}_{\s} \bessel \xistella - \frac{\beta_r}{(k_{\perp} \rho_r)^2} Z_{\s} \tilde{n}_{\s} \vthnorm\bessel \tilde{v}_{\parallel} \zetastella + 2\beta_r \tilde{T}_{\s} \tilde{\mu}_{\s} \frac{J_{1,\s}}{\tilde{a}_{\s}} \sigstella - \hat{C}_{\s}[\lamspec] & = 0 \; ,
	\end{aligned}
	\label{eq:final_lam_eqn_EM_stella}
	\\
	& \xistella  + \frac{1}{\tilde{\eta}} \sum_{\s} \velspace \int \rmd^2 \tilde{v} \left[i \tilde{\omega}_{*,\s} + \zt \tilde{\gamma}^* \right] \bessel \maxwexp \steltildlam =0 \; ,
	\label{eq:final_xi_eqn_EM_stella}
	\\
	&\zetastella -  \sum_{\s} \velspace \int \rmd^2 \tilde{v} \; (2 \vthnorm \tilde{v}_{\parallel})  \left[ i \tilde{\omega}_{*,\s} + \zt \tilde{\gamma}^* \right] \bessel \maxwexp \steltildlam = 0 \; ,
	\label{eq:final_zeta_eqn_EM_stella}
	\\
	&\sigstella -  \sum_{\s} \velspace \int \rmd^2 \tilde{v} \; \left( 4\tilde{\mu}_{\s} \frac{\tilde{T}_{\s}}{Z_{\s}} \frac{J_{1,\s}}{\tilde{a}_{\s}} \right) \left[ i \tilde{\omega}_{*,\s} + \zt \tilde{\gamma}^* \right] \maxwexp \steltildlam = 0 \; ,
	\label{eq:final_sigma_eqn_EM_stella}
\end{align}

\noindent 
with $\steltildlam = \lamstella(\tilde{z},-\tilde{v}_{\parallel},\tilde{\mu}_{\s})$, and $ \tilde{\eta} = \sum_{\s} Z_{\s}^2 \tilde{n}_{\s}/ \tilde{T}_{\s} $. An artificial time dependence has been introduced in equation \eqref{eq:final_lam_eqn_EM_stella} in order to make computation easier, and we solve for the steady state solution for $\lamstella$, when $\partial_{\tilde{t}} \lamstella = 0$. 

As before, we can use ${\hat{G}_{\s} [\bp; \gspec, \tilphi, \tila, \tilb]= \tilde{\gamma} \gspec + \hat{L}_{\s} [\bp; \gspec, \tilphi, \tila, \tilb]}$, with the added constraint of~${\left. \grad_{\bp} \mathcal{L} \right|_{\bp_0} = 0}$ to rearrange the above and obtain:

\begin{align}
	\gradp \tilde{\gamma} \; \langle \gspec, \lamstella \rangle_{\tilde{z},\tilde{v}, {\s}} = - \left. \left[ \langle \partial_{\bp} \hat{L}_{\s},  \lamstella \rangle_{\tilde{z},\tilde{v}, {\s}} + \langle \partial_{\bp} \hat{Q}, \xistella \rangle_{\tilde{z}}  
	+ \langle \partial_{\bp} \hat{M} , \zetastella \rangle_{\tilde{z}} + \langle \partial_{\bp} \hat{N} , \sigstella \rangle_{\tilde{z}} \right] \right|_{\bp_0} \; .
	\label{eq:final_gam_eqn_EM_stella}
\end{align}

\noindent
with closure equations provided by \eqref{eq:final_lam_eqn_EM_stella}-\eqref{eq:final_sigma_eqn_EM_stella}.

\subsubsection{Electrostatic collisionless Limit}
Finally, in the electrostatic, collisionless regime, the system of equations to solve in the normalised \texttt{stella} coordinates is given by

\begin{align}
	\!\begin{aligned}[b]
		\pdv{\steltildlam}{\tilde{t}} + \tilde{\gamma}^* \steltildlam + \gradpar \pdv{\steltildlam}{\tilde{z}}
		- \dvpacoeff \pdv{\steltildlam}{v_{\parallel}} - i \tilde{\omega}_{d, \s} \steltildlam &  \\
		+ Z_{\s} \tilde{n}_{\s} \bessel \maxwexp \xi & = 0 \; ,
		\label{eq:final_lam_eqn_electrostatic_stella}
	\end{aligned}
	&
	\\
	\xi  + \frac{1}{\tilde{\eta}} \sum_{\s} \velspace \int \rmd^2 \hat{v} \left[i \tilde{\omega}_{*,\s} + \zt \tilde{\gamma}^* \right] \bessel \maxwexp \steltildlam = 0 & \; ,
	\label{eq:final_xi_eqn_electrostatic_stella}
\end{align}
\noindent
and
\begin{align}
	\gradp \tilde{\gamma} \; \langle \gspec, \lamstella \rangle_{\tilde{z},\tilde{v}, {\s}} = & - \left. \left[ \langle \partial_{\bp} \hat{L}_{\s}, \lamstella \rangle_{\tilde{z},\tilde{v}, {\s}}  + \langle \partial_{\bp} \hat{Q}, \xistella \rangle_{\tilde{z}} \right] \right|_{\bp_0} .
	\label{eq:electrostatic_derivative_stella}
\end{align}

\subsection{Magnetic Geometry} \label{sec:adjoint_miller}
\noindent
The coefficients in equations \eqref{eq:steady_gyro_eqn_EM_stella}-\eqref{eq:steady_perp_amp_EM_stella}, and thus the associated linear growth rates, are implicitly dependent on the magnetic geometry. For the remainder of the paper we will take $\bp$ to be an appropriate set of parameters that specifies the local magnetic geometry. In particular, we will use the Miller formalism \citep{Miller98} to parameterise the magnetic field on the flux surface of interest. The Miller approach ensures that the Grad-Shafranov \citep{Shafranov66} equation is always satisfied locally by using a set of independent parameters to define a single flux surface in an axisymmetric device. The model equations describing the shape of the fux surface with flux label $r$ are
\begin{align}
	R(r,\theta) = & \; R_0 (r) + r \cos \left\{ \theta + \sin(\theta) \delta(r) \right\} 
	\label{eq:r_coordinate}\\
	Z(r,\theta) = & \; r \; \kappa (r) \sin (\theta).
	\label{eq:z_coordinate}
\end{align}
\noindent 
$R(r,\theta)$ and $Z(r,\theta)$ define the major radial and vertical locations for a given poloidal location, $\theta$, which is related to the cylindrical angle, and $\delta$ and $\kappa$ indicate the triangularity and elongation of the surface respectively. Specifying the full set of Miller parameters provides all of the information required to compute the geometric coefficients in the gyrokinetic-adjoint system consistent with the local MHD equilibrium (though consistency with a global MHD equilibrium is not guaranteed).  


The user specified input parameters used in the local version of \verb|stella| are $\{ r_{\psi_0}, R_{\psi_0}, \Delta_{\psi_0}, q_{\psi_0}, \hat{s}_{\psi_0}, \kappa_{\psi_0}, \kappa'_{\psi_0}, I_{\psi_0}, \delta_{\psi_0}, \delta'_{\psi_0}, \beta'_{\psi_0} \} $, corresponding to markers for the minor and major radii, horizontal Shafranov shift ($\Delta_{\psi_0} = R_{\psi_0}'$ ), safety factor, magnetic shear [$\hat{s} \doteq (r/q) q'$], elongation and its radial derivative, the external axial discharge current (which is used as a proxy for the reference magnetic field), triangularity and its radial derivative, and the radial pressure derivative respectively, with each being specified at the flux surface of interest. Further information about how the Miller geometry is treated in \verb|stella| is given in \apref{sec:geometry}. We henceforth set $\bp \coloneqq \{ r_{\psi_0}, R_{\psi_0}, \Delta_{\psi_0}, q_{\psi_0}, \hat{s}_{\psi_0}, \kappa_{\psi_0}, \kappa'_{\psi_0}, I_{\psi_0}, \delta_{\psi_0}, \delta'_{\psi_0}, \beta'_{\psi_0} \} $. 


\section{Numerical Implementation} \label{sec:numerical_implementation}
\noindent
The aim is to find the magnetic geometry that minimises the linear growth rate for the ITG instability, and maximises the linear critical temperature gradient across the device. This requires three distinct stages: first, the computation of $\gradp \gamma$ at a fixed ion temperature gradient, $T'_{i}$; second, its subsequent use in an optimisation algorithm to find the $\bp$ that minimises $\gamma$ for this given $T'_{i}$; third, iteration of this procedure with variable $T'_{i}$ to find the maximum temperature gradient for which $\gamma \leq 0$ for the range of $\bp$ considered.

\subsection{Initial Simulation} \label{sec:inital_simualation}
\noindent
Solving, at an initial set of $\bp_0$, for $\tilde{\gamma}, \gspec$ and $\tilphi$ is achieved by allowing \texttt{stella} to run for a sufficiently long time, such that the solution is dominated by a single normal mode. To determine the time at which this is satisfied, we employ a convergence test: the growth rate is calculated at each time step and if the value of this is constant in time (within a specified tolerance) then the system is deemed to be converged. There are two components of the convergence test. The first is to check that the growth rates calculated at adjacent time steps are within a given tolerance of each other. The second is to perform a windowed average to check that the growth rate remains consistent over a defined number of time steps. Two windowed averages are done; one over $N_t$ time steps and one over $\mathbb{Z}(N_t/2) $ time steps. When these two window-averaged growth rates agree within a set tolerance, $\gspec$ and $\tilphi$ are taken to have converged. The corresponding growth rate is then calculated from the windowed average.

\subsection{Adjoint Simulation} \label{eq:Adjoint_simulation}
\noindent
As previously introduced, an artificial time dependence is added to the adjoint equations to facilitate computation. The solution is found in the steady state limit, in which the time derivative appearing in the adjoint equations goes to zero. 

The resulting adjoint equations are treated in a similar way to the treatment of the usual gyrokinetic system of equations in \texttt{stella}. The main aim is to ensure that the parallel streaming term may be treated separately from the rest of the dynamics through operator splitting. This is done by discretising in time, and splitting the time derivative into a series of three steps:
\begin{align}
	\pdv{\steltildlam}{\tilde{t}} = \left(\pdv{\steltildlam}{\tilde{t}}\right)_1 + 	\left(\pdv{\steltildlam}{\tilde{t}}\right)_2 + 	\left(\pdv{\steltildlam}{\tilde{t}} \right)_3 ,
\end{align}
\noindent
where \addtocounter{equation}{-1}
\begin{subequations}
\begin{align}
	\left(\pdv{\steltildlam}{\tilde{t}}\right)_1 = & - \tilde{\gamma}^* \steltildlam + i \tilde{\omega}_{d, \s} \steltildlam - Z_{\s} \tilde{n}_{\s} \xistella ,
	\label{eq:explicit_lam_eqns} \\
	\left(\pdv{\steltildlam}{\tilde{t}}\right)_2 = & \; \dvpacoeff \pdv{\steltildlam}{\tilde{v}_{\parallel}} ,
	\label{eq:mirror_lam_eqns} \\
	\left(\pdv{\steltildlam}{\tilde{t}} \right)_3 = & - \gradpar \pdv{\steltildlam}{\tilde{z}} .
	\label{eq:implicit_lam_eqns}
\end{align}
\end{subequations}
%
%
%
%
Analogous to \verb|stella|, the terms in \eqref{eq:explicit_lam_eqns} are treated explicitly using a strong-stabiity-preserving, third order Runge-Kutta method (\citealt{Gottlieb01}). The parallel streaming and mirror terms, given by equations \eqref{eq:implicit_lam_eqns} and \eqref{eq:mirror_lam_eqns} respectively, are treated separately, due to the presence of the prefactor $\vthnorm$, which increases the relative amplitude of these terms when considering electron dynamics. As a result these terms have the potential to exert a stringent Courant–Friedrichs–Lewy (CFL) condition on the simulation, and require a small time step to be taken in order to retain accuracy. Thus these terms are treated implicitly in time to relax this constraint. 

The mirror term, \eqref{eq:mirror_lam_eqns}, is an advection equation of $\steltildlam$ in $\tilde{v}_{\parallel}$, which is treated using a semi-Lagrange method, akin to the algorithm used to advect the distribution function in $\tilde{v}_{\parallel}$ in \texttt{stella}. The streaming term, \eqref{eq:implicit_lam_eqns}, is also an advection equation in $\tilde{z}$, which is treated using the Thompson algorithm for tri-diagonal solve. 

We are seeking a steady state solution to the adjoint equations. The same convergence test is performed on $\steltildlam$ as that performed on the distribution function in order to check that the complex growth rate has converged to zero within a given tolerance. When this is satisfied, the resulting $\steltildlam$ is used to solve for $\lamstella$. This is then stored for use in the remainder of the calculation. 


\subsection{Optimisation Loop}\label{eq:optimisation_loop}
\noindent
Once the gradients $\gradp \gamma$ are obtained we use them inside an optimisation loop to find the $\bp$ that minimises $\gamma$. We employ the Levenberg-Marquardt (LM) algorithm, \citep{Transtrum12}, to find the local minimum. The method adopts a steepest descent behaviour when the location in parameter space is considered to be far from the minimum, and progresses towards Gauss-Newton behaviour as one approaches the minimum. This is achieved by introducing a damping factor, $\Gamma$, which is updated with each iteration. The algorithm iteratively solves the following:

\begin{align}
	[\vec{H} + \Gamma \; \text{diag}(\vec{H}) ] \rmd \bp = - \gradp \gamma ,
	\label{eq:LM_algorithm}
\end{align}

\noindent
with $\vec{H} = \gradp^2 \gamma \approx (\gradp \gamma)^{\dag} \gradp \gamma$ the Hessian matrix. When $\Gamma$ is large we have $\bp_{\text{new}} \approx \bp_{\text{old}} - \vec{\alpha}\cdot \grad_{\bp} \gamma$, with $\vec{\alpha} \doteq [\Gamma \; \text{diag} (\vec{H})]^{-1}$, which mimics the gradient descent algorithm. However, when $\Gamma$ is small \eqref{eq:LM_algorithm} reduces to $\bp_{\text{new}} \approx \bp_{\text{old}}  - \vec{H}^{\dag} \cdot \gradp \gamma$, which matches with the Gauss-Newton algorithm. 

The LM formalism is derived using the Taylor expansion, and as such a trust region is included within the optimisation loop to ensure that the updated value of $\bp$ is close enough to the previous, such that the Taylor approximation is valid within the limits for which the algorithm is applied. The trust region for $\bp$ is defined via:

\begin{align}
	\rho = \frac{0.5 \;\rmd \bp^{\dag}  \cdot \vec{H} \cdot \rmd \bp }{\rmd \bp \cdot \gradp \gamma} < \bar{\epsilon} ,
\end{align}

\noindent
where $\bar{\epsilon}$ is a chosen tolerance. If $\rho > \epsilon$ for a given $\rmd \bp$ the algorithm rejects the output $p_{\text{new}}$ and increases the weight $\Gamma$ in an attempt to improve the accuracy of the approximation. This helps ensure that the updated value of $\bp$ is a reasonable one. 

It is worth noting that the LM algorithm is designed to find local minima, so there is no guarantee that the minimum obtained is the global minimum of the system. 

We also emphasise that the gradient-based optimisation algorithm is independent of the adjoint method that has been developed for gyrokinetic microstability. The optimisation loop may be itself optimised to efficiently search for regions of stability given a gradient input. Different alorithms, and indeed different parameter choices within each algorithm, will yield different efficiencies in finding stable solutions. An illustrative example of this is given later in Figure \ref{fig:plot_for_kap_tri_gam} where the step size for the optimisation loop is varied to yield two distinct paths through the parameter space using the adjoint gradient. However, this is not the focus of this paper and we will not labour on enhancing the gradient-based optimisation loop.


\section{Numerical Results}\label{sec:Numerical_results}

\subsection{Initial Benchmark}\label{sec:benchmark_oli}
\noindent
The first numerical check we perform is to ensure that the values of $\rmd_{\bp} \gamma_0$ obtained from our adjoint method agree with those obtained using a finite difference approach. 
Following this we perform a more extensive benchmark by conducting a parameter scan in the growth rate using \texttt{stella} for different values of the Miller parameters. We then choose an initial set of parameters and perform the adjoint-optimisation scheme described above, and overlay the results of this with the parameter scan to see how these compare, whilst checking the growth rates at each point considered by the adjoint-optimisation scheme against those obtained using a finite-difference approach. As a proof of principle we chose to vary two parameters: triangularity, $\delta$, and elongation, $\kappa$, whilst holding the other Miller variables fixed. Given that the Miller parametrisation is local to a given flux surface, this variation is not necessarily consistent with a global solution to the Grad-Shafranov equation. The choice to vary these two parameters in isolation is driven by two considerations; first, it is only intended as a proof-of-principle check of the adjoint approach so simplicity is desirable, and second, previous research has shown that maximal shaping, with large elongation and triangularity, minimises the linear ITG instability, whereas parameters such as $\kappa'$ and $\delta'$ have an order $a/R_0 \ll 1$ is the inverse aspect ratio, with $a$ and $R_0$ the scales associated with the minor and major radii respectively, \cite{Highcock18}. Table \ref{tab:equilibrium_miller} lists the values of input equilibrium variables in the Miller geometry. These have been chosen to coincide with values used in \cite{Beeke20} in order to verify the qualitative behaviour found.

Given these initial values of $\delta$ and $\kappa$, we determine the linear growth rates for a grid of perpendicular wavenumbers within $k_x<2.0$, $k_y<2.0$, which reveals that the most unstable mode is found at $\{k_y, k_x\} = \{0.68, 0.0\}$ for mass ratio $m_{e}/m_{i} = 2.7 \times 10^{-4}$, normalised species temperatures and densities of $T_i = T_e = 1$, $n_i = n_e = 1$, and normalised species temperature and density gradients of $a/L_{T_i} = a/L_{T_e} = 2.42$, $a/L_{n_i} = a/L_{n_e} = 0.81$.

\begin{table}
	\begin{center} \setlength\tabcolsep{14.0pt} \caption{List of Miller Parameters} 	
		\label{tab:equilibrium_miller}
		\begin{threeparttable}\renewcommand{\arraystretch}{1.0}\footnotesize 
			\begin{tabular}{ m{6em}  m{2cm} } \toprule\midrule
				\emph{Miller} & \emph{Initial} \\
				\emph{Parameter } & \emph{Value}
				\\ \midrule 
				$\rho$ & 0.5 \\ 
				$R_0$ & 2.94 \\
				$R_{geo}$ & 2.94 \\
				$\Delta$ & -0.11 \\
				$\kappa$ & 1.52 \\
				$\kappa'$ & 0.10\\
				$q$ & 2.02 \\
				$\hat{s}$ & 0.34 \\
				$\delta$ & 0.14 \\
				$\delta'$ & 0.29 \\ 
				$\beta'$ & 0.069 \\ 
				\midrule \bottomrule
			\end{tabular} 
			\begin{tablenotes}[flushleft] \item Equilibrium Miller parameter values used in the initial benchmark simulations.
        \end{tablenotes} 
		\end{threeparttable} 
	\end{center} 
\end{table}
\begin{figure}
	\centering
	\includegraphics[scale=1.0]{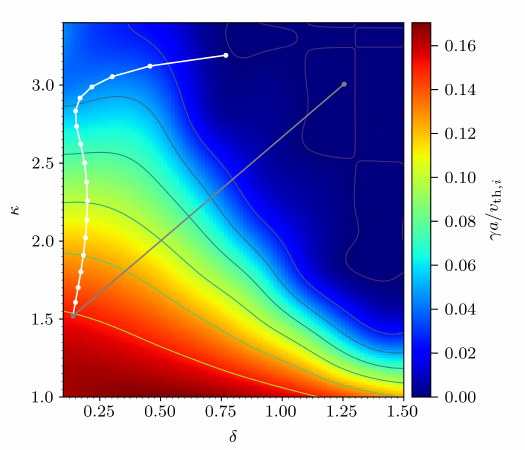}
	\caption{Two-dimensional parameter scan over elongation and triangularity, with the colour map indicating the amplitude of the linear growth rate. Here $k_y = 0.68$, $k_x = 0.0$, $m_{e}/m_{i} = 2.7 \times 10^{-4}$, $T_i = T_e = 1$, $n_i = n_e = 1$, $a/L_{T_i} = a/L_{T_e} = 2.42$, $a/L_{n_i} = a/L_{n_e} = 0.81$, with $a$ the minor radius of the last closed flux surface. The path taken by the optimisation algorithm is indicated in white, with the initial point $\kappa = 1.5$, and $\delta= 0.14$. A second path, drawn in grey, is shown indicating the adjoint optimisation with a different step size for the optimisation loop.}
	\label{fig:plot_for_kap_tri_gam}
\end{figure}

In Figure~\ref{fig:plot_for_kap_tri_gam} we show a scan in the linear growth in elongation and triangularity obtained by running \texttt{stella} with the Miller parameters specified in Table \ref{tab:equilibrium_miller}, and the values of $\kappa$ and $\delta$ adjusted accordingly. The contour colour indicates the magnitude of the growth rate, and the plot extends over a range of values that has been set by reasonable physical constraints on devices.

Figure \ref{fig:plot_for_kap_tri_gam} shows that increasing the elongation and triangularity of our flux surface reduces the linear growth rate, and that there exists a region of stability when the shaping is maximal, in agreement with previous work. The path taken using the adjoint method is indicated in white. At the chosen starting point, located in the region of instability, the gradient $\rmd_{\bp}\gamma$ is calculated and the value of $\bp = \{\delta, \kappa\}$ is updated using the previously mentioned LM method. The final point is found to be locally stable as the growth rate here is negative. The algorithm then checks a nearby point to determine if a small region of stability exists and, once this has been verified, outputs this as the final $\bp$ value. In this particular case of a 2D parameter scan, a finite difference approach could utilise the same  optimisation loop, requiring only three simulations, plus one additional simulation performed at the next iteration point to verify the growth rate. The adjoint method must simulate the gyrokinetic equation once, and also solve the adjoint equations -- which are of similar computational cost as the gyrokinetic equation, in order to achieve the same result. In this particular low-dimensionality case, the advantage of the adjoint approach is comparatively small over using a finite difference approach. However, the computation time required for the traditional method scales linearly with the number of parameters, so that for an $N$-dimensional parameter space, $N+1$ simulations are necessary at each iteration point, whilst the adjoint method is effectively independent of the number of parameters, and only the steady state solutions to the gyrokinetic and adjoint equations are required. It can therefore be readily applied to a high-dimensional parameter space without any significant increase in computational cost. 

A second path is plotted on Figure~\ref{fig:plot_for_kap_tri_gam} in grey. This path is taken using the same adjoint technique, but increasing the step size within the optimisation loop. When the step-size is small, as with the white path, the LM algorithm more closely resembles a gradient-descent method, however when the step size is larger, as with the grey path, the LM algorithm resembles Newton's method for gradient optimisation. The figure illustrates how the adjoint algorithm can be combined with an optimiser to quickly and efficiently converge to a stable region of parameter space. 
\subsection{Increasing the Critical Temperature Gradient}
\noindent
We have seen that the adjoint method is a powerful technique for computing stable points in a large parameter space; Figure \ref{fig:plot_for_kap_tri_gam} shows that it can efficiently be used inside an optimisation loop to find a minimum of the linear growth rate for a given temperature gradient. Once the adjoint optimisation loop locates a region in the parameter space with negative or zero growth rate the local temperature gradient (or other plasma profile variable of interest) may be increased and the process repeated. Conversely, if a minimum positive (unstable) growth rate is found, the temperature gradient can be reduced to seek out the optimal shape that maximises the critical temperature gradient at which linear instability occurs to seek out microstability.

\begin{figure}
	\centering
	\includegraphics[scale=1.0]{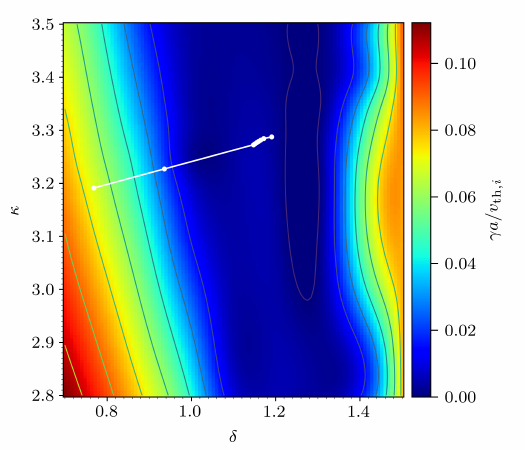}
	\includegraphics[scale=1.0]{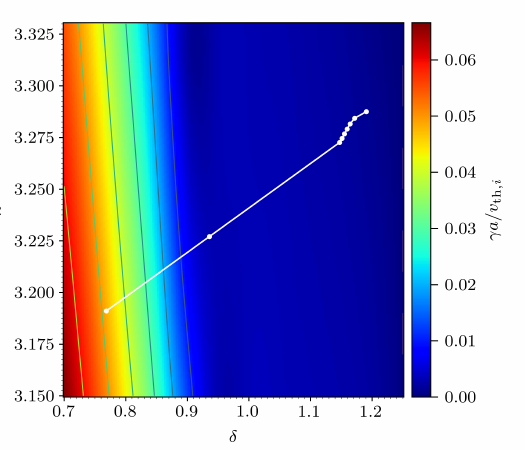}
	\caption{Plots showing a parameter scan in elongation and triangularity, with a temperature gradient of $a/L_{T_i} = 3.80$, increased from the $a/L_{T_i} = 2.42$ value used in Figure \ref{fig:plot_for_kap_tri_gam}. The geometry of the initial point, located in the unstable region, is provided by final point in Figure \ref{fig:plot_for_kap_tri_gam} and is now unstable due to the increased temperature gradient. Here $k_y = 0.68$, $k_x = 0.0$, $m_{e}/m_{i} = 2.7 \times 10^{-4}$, $T_i = T_e = 1$, $n_i = n_e = 1$, $a/L_{T_i} = a/L_{T_e} = 3.80$, $a/L_{n_i} = a/L_{n_e} = 0.81$. Note that the colour scales used in the figures above are different than that used in Figure \ref{fig:plot_for_kap_tri_gam}. The right hand side plot is a zoomed in figure of the left.}
	\label{fig:increased_temp}
\end{figure}
\FloatBarrier

Figure \ref{fig:increased_temp} demonstrates an iterative use of the adjoint optimisation. Here we have iterated the temperature gradient, and inside each temperature gradient iteration the adjoint method is used to find a stable geometric configuration. However this principle could be extended further by continuing to increase the temperature gradient until no region of stability is available, indicating a limiting temperature gradient that can be achieved through geometric considerations alone. Figure \ref{fig:increased_temp} is a demonstration that the adjoint method may be used to increase the temperature gradient, whilst retaining stability using geometry. Though we have opted to consider only two parameters in the above as a demonstration and for clarity have focused on ITG, it is possible to employ the adjoint method to optimise over a large number of geometric parameters simultaneously. Such an exploration would be expensive using traditional finite difference methods. 
\subsection{Negative Triangularity}
\noindent
To form a final example here, we note there has been previous evidence that negative triangularity can offer improvements for microstability, \cite{Pochelon99}, so we repeated the benchmark scan using negative triangularity. All values of equilibrium Miller parameters are the same as in Table \ref{tab:equilibrium_miller}, except we now set the initial value of triangularity to $\delta = -0.14$.

\begin{figure}
	\centering
	\includegraphics[scale=1.0]{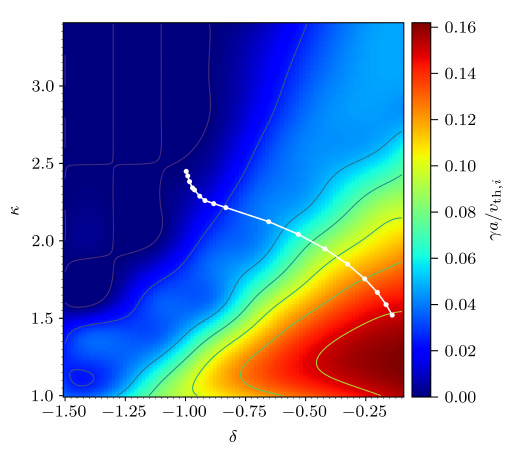}
	\caption{Growth rate contours for a parameter scan with negative triangularity for $k_y = 0.68$, $k_x = 0.0$ and equilibrium parameters $m_{e}/m_{i} = 2.7 \times 10^{-4}$, $T_i = T_e = 1$, $n_i = n_e = 1$, $a/L_{T_i} = a/L_{T_e} = 3.80$, $a/L_{n_i} = a/L_{n_e} = 0.81$. The white line indicates the path taken by the optimisation algorithm. The initial values of $\{\delta, \kappa\}$ are taken to be $\{-0.14, 1.52\}$.}
	\label{fig:negative_triangularity}
\end{figure}
\FloatBarrier

\noindent
The path taken using the adjoint optimisation loop is again shown in Figure \ref{fig:negative_triangularity} by the white path, starting in the dark red red region at $\{\delta, \kappa\} = \{-0.14, 1.52\}$. 

\begin{figure}
	\centering
	\includegraphics[width=100mm,height=60mm]{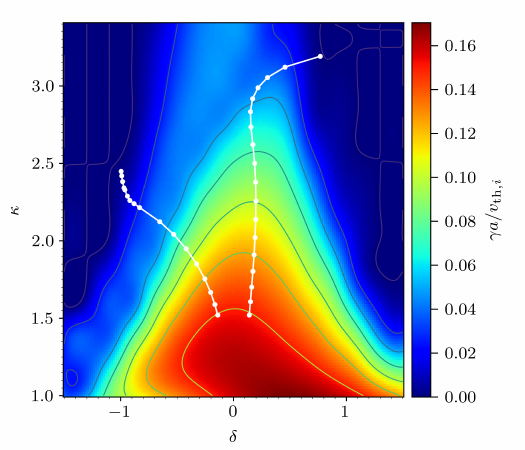}
	\caption{Growth rate contours for a parameter scan with both positive and negative triangularity for $k_y = 0.68$, $k_x = 0.0$ and equilibrium parameters $m_{e}/m_{i} = 2.7 \times 10^{-4}$, $T_i = T_e = 1$, $n_i = n_e = 1$, $a/L_{T_i} = a/L_{T_e} = 3.80$, $a/L_{n_i} = a/L_{n_e} = 0.81$. The white line indicates the two paths taken by the optimisation algorithm starting in different regions in parameter space. The initial values of $\{\delta, \kappa\}$ are taken to be $\{0.14, 1.52\}$ and $\{-0.14, 1.52\}$.}
	\label{fig:pos_neg_triangularity}
\end{figure}

This highlights a key feature of the gradient-based optimisation method: the solution is not unique, and the output can depend on the starting region within parameter space. See Figure \ref{fig:pos_neg_triangularity} for an illustrative example of this.

\FloatBarrier

\begin{figure}
	\centering
	\includegraphics[scale=0.7]{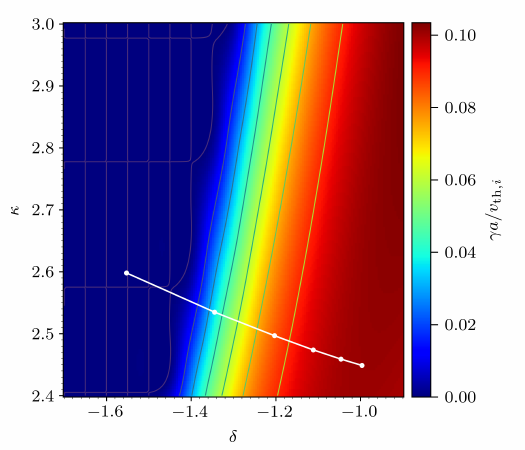}
	\caption{Growth rate contours for a parameter scan with negative triangularity at a temperture gradient of $a/L_{T_i}= 3.80$, increased from $a/L_{T_i}= 2.42$. The geometry of the initial point is taken as the final point in Figure~\ref{fig:negative_triangularity}, and is now unstable due to the increased temperature gradient. The scan is performed at the same parameter values as those in Figire~\ref{fig:negative_triangularity} -- $k_y = 0.68$, $k_x = 0.0$ $m_{e}/m_{i} = 2.7 \times 10^{-4}$, $T_i = T_e = 1$, $n_i = n_e = 1$, $a/L_{T_i} = a/L_{T_e} = 3.80$, $a/L_{n_i} = a/L_{n_e} = 0.81$. The white line indicates the path taken by the optimisation algorithm. The initial values of $\{\delta, \kappa\}$ are taken to be $\{-0.9965, 2.4488\}$.}
	\label{fig:negative_triangularity_high_temp}
\end{figure}

Finally, we consider increasing the temperature gradient for the negative triangularity case, and we repeat the procedure to find a stable region of parameter space. We again take the input parameters $\{\delta, \kappa \} = \{-0.9965, 2.4488\}$ to be the outputs from the previously optimised case at a temperature gradient of $a/L_{T_i} = 2.42$. Then we use the adjoint algorithm coupled with the gradient optimiser to look for a stable region of parameter space at an increased temperature gradient of $a/L_{T_i} = 3.8$. The results of this are shown in Figure \ref{fig:negative_triangularity_high_temp}. 

\FloatBarrier
\section{Numerical Efficiency Improvement}

Recall that a conventional finite difference approach method for a parameter vector, $\bp$, of size $\mathcal{N}$ we require $\mathcal{N}+1$ simulations to be carried out until convergence at each of the points considered (these are shown by the white dots in the figures above) in order to compute one gradient in our parameter space. However, when using the adjoint method the same gradient may be computed for the cost of roughly two simulations - one gyrokinetic, and one adjoint, which is of similar computational cost as a gyrokinetic simulation. This approach is essentially independent of the number of parameters used, with the only additional cost incurred being that associated with calculating the partial derivatives that appear. However, these are extremely inexpensive for computers and can be calculated using one processor.

To illustrate this example we will show the progressive computational improvement of using the adjoint method, as compared with finite differences, and show how this scales favourably with increasing $\mathcal{N}$. In order for the comparisons to be fair, we shall run all simulations to a standard time of $100 a/v_{\mathrm{th}, i} $.

\subsection{Numerical demonstration of improved efficiency}
\noindent
For the case where $\mathcal{N} =2$ we are optimising with respect to two parameters, $\{ \delta, \kappa\}$. We find that to calculate a gradient at each point in our parameter space requires a total of $\sim 4.752$ CPU (Central Processing Unit) hours. The same calculation as performed using the adjoint method requires a total of $\sim 3.168$. Though this is only a modest improvement, we shall show that as $\mathcal{N}$ increases the advantage of using the adjoint method, over a finite difference scheme, becomes increasingly apparent. 

Increasing the number of parameters to $\mathcal{N} = 4$, optimising over the miller parameters $\{ \Delta, \kappa, q,\delta \}$, we find that the finite difference approach requires $\sim 7.920$ CPU hours to compute a gradient at each point in parameter space. However, when computing the same gradient using the adjoint method the CPU time, to the precision of the CPU clock, is $\sim3.168$ CPU hours.

If we increase the number of parameters further to $\mathcal{N} = 7$, optimising over the miller parameters $\{R_0, R_{\mathrm{geo}}, \Delta, \kappa, q, \hat{s}, \delta \}$ we find that using the finite difference approach requires $\sim 12.672$ CPU hours to compute each gradient at a point in parameter space. When using the adjoint method to compute the same  gradient, to the precision of the CPU clock, is still $\sim 3.168$ CPU hours.

Hence, we conclude that the numerical speed up is significant with increasing $\mathcal{N}$. This allows for the potential of including multiple harmonics within our shaping optimisation for very little additional computing cost. 

\begin{figure}
	\centering
    \includegraphics[scale=0.40]{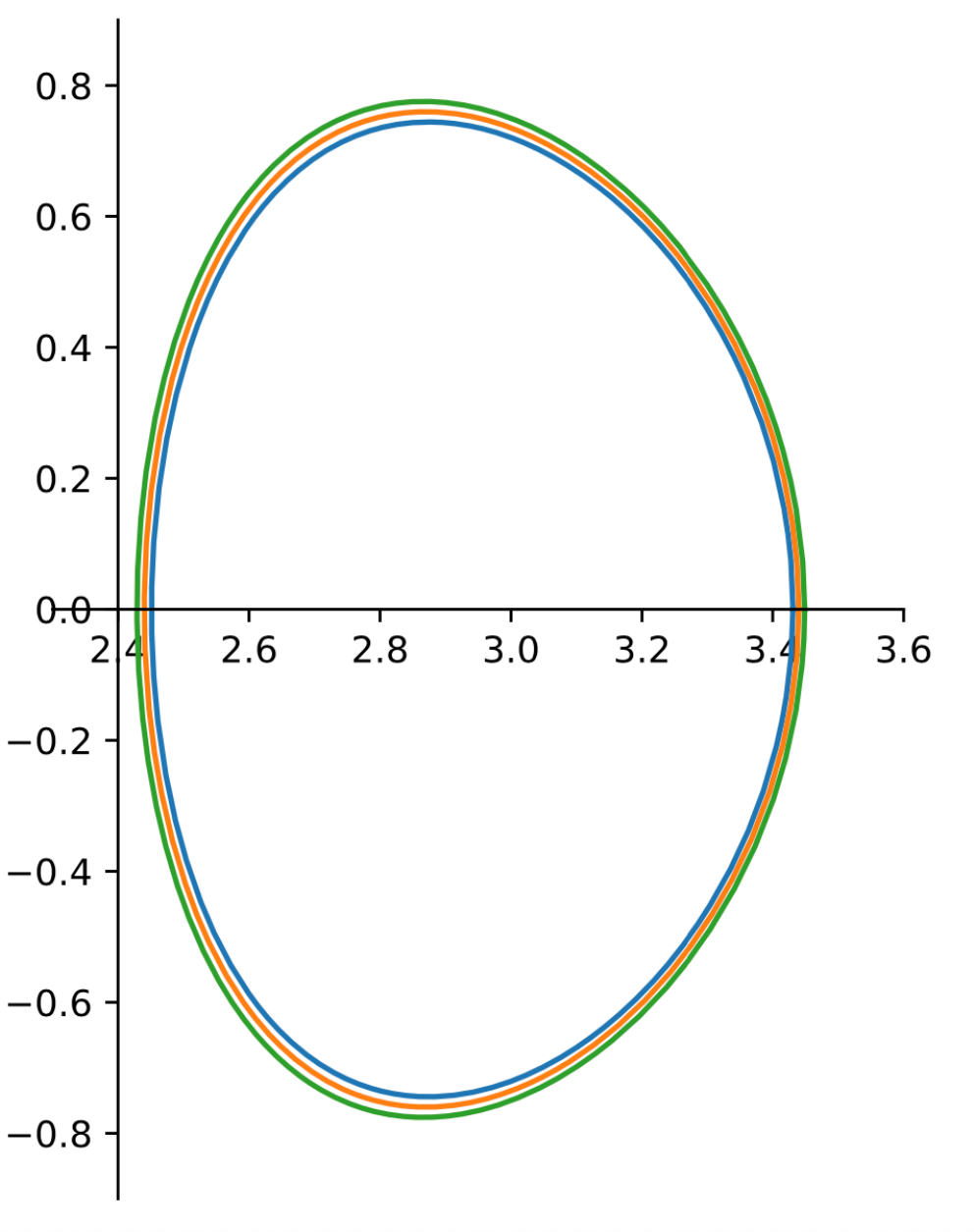}
	\includegraphics[scale=0.40]{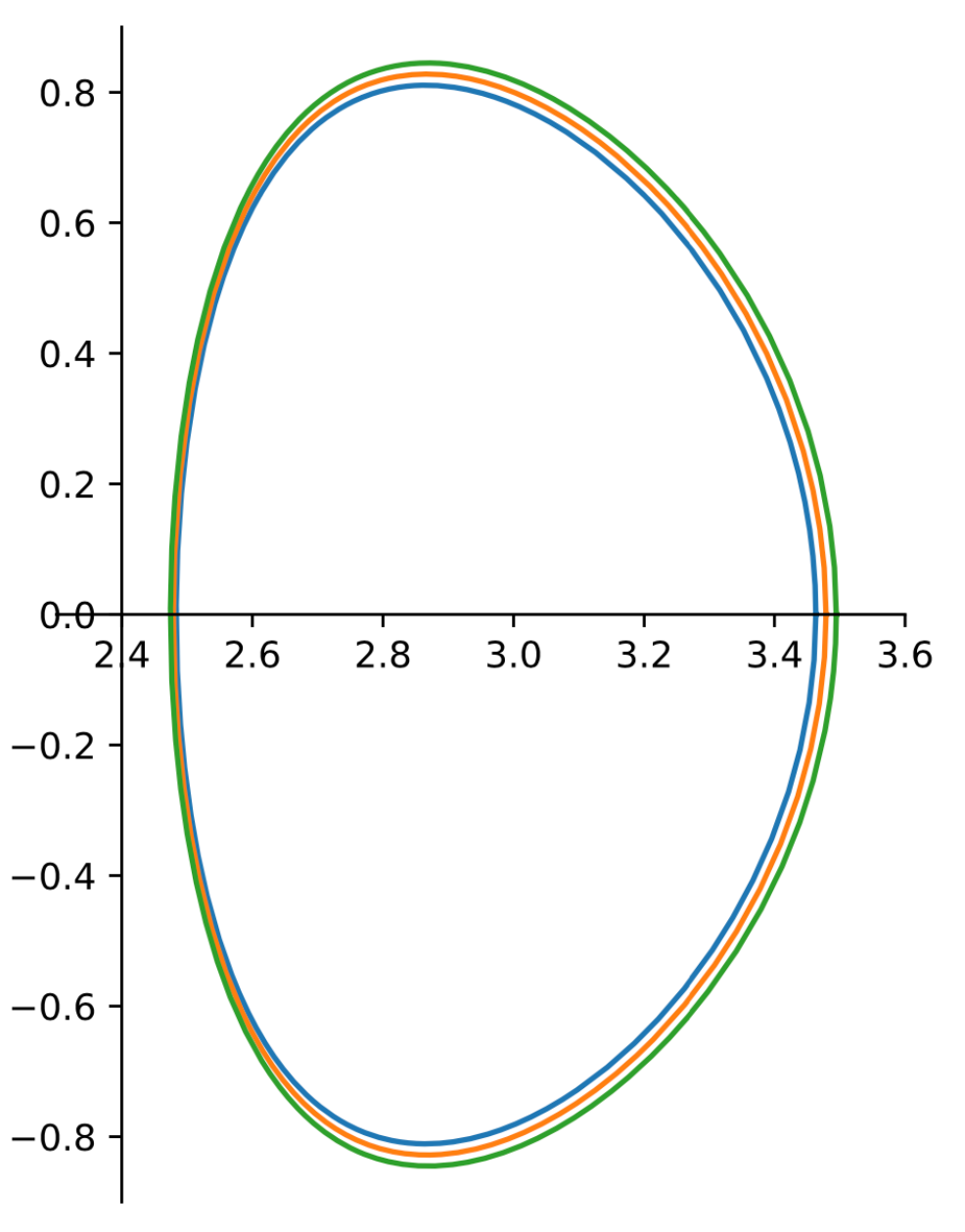}
	\caption{Plots of the flux surfaces in the poloiodal cross section. The orange surface is the flux surface at $\rho = 0.5$, and the blue and green surfaces are the two adjacent flux surfaces. The image on the left is the initial unstable configuration, before optimising. The image on the right is the stable, optimised configuration found using the adjoint-LM system. }
	\label{fig:Miller_plots}
\end{figure}

For the example including seven parameters, we iterate the process of stepping through parameter space to find a point of stability. We take our set of initial parameters to be those given in Table~\ref{tab:equilibrium_miller}, and we simulate using a temperature and density gradient of $a/L_{T_i} = a/L_{T_e} = 2.42$ and $a/L_{n_i} = a/L_{n_e} = 0.81$ respectively. We perturb the magnetic geometry by varying $\{R_0, R_{\mathrm{geo}}, \Delta, \kappa, q, \hat{s}, \delta \}$. Following iterations of the coupled adjoint-LM system we find a configuration that is stable when $\{R_0, I, \Delta, \kappa, q, \hat{s}, \delta \} = \{2.979, 2.846, 0.562, 1.656, 2.085, 0.167, 0.225\}$. A cross section of the initial, unstable configuration and the final, stable configuration, with their corresponding neighbouring flux surfaces, are shown in Figure~\ref{fig:Miller_plots}.

\FloatBarrier

\section{Conclusion and Discussion} \label{sec:conclusion}
\noindent
We have derived an adjoint method tailored for local, linear gyrokinetics and elucidated its numerical integration within the $\delta \! f$-gyrokinetic code \texttt{stella}. 
As a proof-of-principle case we have demonstrated the effectiveness of our adjoint method, as applied to a
gyrokinetic system, by finding the geometric configuration of the magnetic field that is stable to microinstabilities. The illustrative example given has focused on increasing the temperature gradient, whilst preserving microstability; ion temperature gradient (ITG) instabilities are often prevalent in fusion devices due to the existence of large temperature gradients, and thus it is conceivable that geometric considerations could help mitigate their growth and improving overall efficiency.

As that the computational cost of the adjoint method remains independent of the number of parameters, its applicability to high-dimensional parameter spaces is readily apparent. The advantages become more pronounced with an expanding number of parameters, as the adjoint method outperforms traditional techniques for calculating gradients, where the computation cost scales with parameter count. This becomes especially beneficial when examining devices like stellarators, which have a large number of geometric parameters that can influence the microinstability of the confined plasma.

It is important to stress that although we have demonstrated a specific example focused on increasing the temperature gradient, this approach can be readily extended to increase the density gradient or other plasma properties, by adapting the overarching LM optimisation loop, without necessitating alterations to the adjoint calculation itself. Such adaptability enables the application of \texttt{stella} and other local, $\delta \! f$-gyrokinetic codes to explore the impact of shaping on various types of microinstabilities and assess how geometry can be instrumental in mitigating their growth.

A crucial point to note is that while the numerical examples presented above have focused on the electrostatic, collisionless regime for optimization with respect to the Miller geometry, equations \eqref{eq:final_lam_eqn_EM}-\eqref{eq:final_gam_eqn_EM} maintain generality. They can be applied to an electromagnetic system, including collisions, and can be optimized using any appropriate set $\{p_i\}$.

\section{Acknowledgement}
This work has been carried out within the framework of the EUROfusion Consortium, funded by the European Union via the Euratom Research and Training Programme (Grant Agreement No 101052200 — EUROfusion)  and from the EPSRC [grant number EP/W006839/1]. Views and opinions expressed are however those of the author(s) only and do not necessarily reflect those of the European Union or the European Commission. Neither the European Union nor the European Commission can be held responsible for them. Part of the simulations presented were performed on the CINECA Marconi supercomputer within the framework of the FUA37\_STELTURB project.
\newpage
\appendix

\section{Decomposition of Operators} \label{sec:decomp_operators}
Here we give the definitions of the operators introduced in \secref{sec:Adjoint_method_GK}  in equations \eqref{eq:GK_decomposition_EM}-\eqref{eq:quasi_decomposition_EM}
\begin{align}
	\hat{G}_{g,\s} [\bp; \gun] = & \gamma  \gun + v_{\parallel} \bdotgradzun \pdv{\gun}{z}- \frac{\mu}{m_{\s}} \bdotgradzun \pdv{B_0}{z} \pdv{\gun}{v_{\parallel}} + i \omega_{d,\s} \gun,
	\nonumber \\
	\hat{G}_{\phi,\s} [\bp; \pun] = & v_{\parallel} \bdotgradzun \frac{Z_{\s} e }{T_{\s}} \maxwellian  \pdv{\bessel \pun }{z}  + i \frac{Z_{\s} e }{T_{\s}}  \omega_{d,\s} \bessel \maxwellian \pun + i \omega_{*,\s} \bessel \maxwellian \pun,
	\nonumber \\
	\hat{G}_{A_{\parallel},\s} [\bp; \aun] = & - \frac{v_{\parallel}^2}{c} \bdotgradzun \frac{Z_{\s} e }{T_{\s}} \maxwellian  \pdv{\bessel \aun }{z} -  i \frac{Z_{\s} e }{T_{\s}}  \omega_{d,\s} \bessel \maxwellian \frac{v_{\parallel} } {c} \aun - i \omega_{*,\s} \bessel \maxwellian \frac{v_{\parallel}}{c} \aun 
	\nonumber \\
	& + \frac{Z_{\s} e}{T_{\s}} \frac{\mu_{\s}}{m_{\s} c } \hat{\boldsymbol{b}} \cdot \grad B_0 \maxwellian \bessel \aun ,
	\nonumber \\ 
	\hat{G}_{B_{\parallel},\s} [\bp; \bun ] = & 2 v_{\parallel} \bdotgradzun \maxwellian \pdv{}{z} \left(  \frac{\mu_{\s}}{T_{\s}} \frac{J_{1,\s} }{a_{\s }} \bun \right) + 2 i \omega_{d,\s}  \frac{\mu_{\s}}{T_{\s}} \frac{J_{1, \s} }{a_{\s }} \maxwellian \bun  
	\nonumber \\ 
	& +  2 i \omega_{*,\s} \frac{\mu_{\s}}{Z_{\s} e} \frac{J_{1, \s} }{a_{\s }} \maxwellian \bun ,
	\nonumber \\ 
	\hat{Q}_{g,\s} [\bp; \gun] = & Z_{\s} e  \bessel \gun,
	\nonumber \\
	\hat{Q}_{\phi} [\bp; \pun] = & \sum_{\s} \frac{(Z_{\s} e)^2 n_{\s}}{T_{\s}} (\Gamma_{0,\s} -1 ) \pun ,
	\nonumber \\
	\hat{Q}_{B_{\parallel}} [\bp; \bun] =  & \sum_{\s} Z_{\s} e n_{\s} \frac{\Gamma_{1,\s}}{B_0} \bun,
	\nonumber \\
	\hat{M}_{g,\s} [\bp; \gun] = & - \frac{4 \pi}{k_{\perp}^2}  \frac{v_{\parallel} }{c}  Z_{\s} e  \bessel \gun,
	\nonumber \\
	\hat{M}_{A_{\parallel}} [\bp; \aun] = &  \left[1 + \frac{4 \pi}{ k_{\perp}^2  c^2} \sum_{\s} \frac{ (Z_{\s} e)^2 n_{\s}}{m_{\s}} \Gamma_{0, \s} \right] \aun ,
	\nonumber \\
	\hat{N}_{g,\s} [\bp; \gun] = & 8 \pi \frac{J_{1,\s} }{a_{\s} } \mu_{\s} \gun ,
	\nonumber \\
	\hat{N}_{\phi} [\bp; \pun] = &   \left[ 4 \pi \sum_{\s} \frac{Z_{\s} e n_{\s}}{B_0} \Gamma_{1, \s} \right] \pun ,
	\nonumber \\ 
	\hat{N}_{B_{\parallel}} [\bp; \bun] = &  \left[1+ 16 \pi \sum_{\s} \frac{ n_{\s} T_{\s}}{B_0^2} \Gamma_{2, \s} \right] \bun, 
\end{align}

\noindent
and $\hat{C}_{\s,\s'}$ is an appropriate collision operator. 

\section{Adjoints of Operators}  \label{sec:adjoint_operators}
The adjoint operators appearing in equations \eqref{eq:lam_symbolic_eq}-\eqref{eq:sigma_symbolic_eq} are obtained by performing integration by parts wherever a derivative acts on the distribution function or a field variable. After performing the change of variables $\tilde{v}_{\parallel} \rightarrow - \tilde{v}_{\parallel}$ the adjoint operators take the form:
\begin{align}
	\hat{G}_{g,\s}^{\dag} [\bp; \tildlam] & = \gamma^* \tildlam + v_{\parallel} \bdotgradzun \pdv{\tildlam}{z}- \frac{\mu_{\s} }{ m_\s} \bdotgradzun \pdv{B_0}{z} \pdv{\tildlam}{v_{\parallel}} - i \omega_{d,\s} \tildlam ,
	\nonumber \\
	\hat{G}_{\phi,\s}^{\dag} [\bp; \tildlam] & =  \frac{Z_{\s} e} {T_{\s} } \bessel \maxwellian \hat{S}_{\s}, 
	\nonumber \\
	\hat{G}_{A_{\parallel},\s}^{\dag} [\bp; \tildlam] & = \frac{Z_{\s} e} {T_{\s} } \bessel \maxwellian \frac{v_{\parallel}}{c } \hat{S}_{\s}
	+ \frac{Z_{\s} e}{T_{\s}} \frac{\mu_{\s}}{m_{\s} c } \hat{\boldsymbol{b}} \cdot \grad B_0 \maxwellian \bessel \tildlam ,
	\nonumber \\
	\hat{G}_{B_{\parallel},\s}^{\dag} [\bp;\tildlam] & = 2 \frac{ \mu_{\s}}{T_{\s} } \frac{J_{1,\s}}{\tilde{a}_{\s}} \maxwellian \hat{S}_{\s}, 
	\nonumber \\
	\hat{Q}_{g,\s}^{\dag} [\bp; \xi] &= Z_{\s} e \bessel \maxwellian \xi,
	\nonumber \\
	\hat{Q}_{\phi}^{\dag} [\bp; \xi] & = \sum_{\s} \frac{(Z_{\s} e)^2 n_{\s}}{T_{\s}} \left(\Gamma_{0,\s} -1 \right) \xi,
	\nonumber \\
	\hat{Q}_{B_{\parallel}}^{\dag} [\bp; \xi] & = 4 \pi \sum_{\s} \frac{Z_{\s} e n_{\s} } {B_0} \Gamma_{1,\s} \xi ,
	\nonumber \\
	\hat{M}_{g,\s}^{\dag} [\bp; \zeta] & = - \frac{4 \pi}{k_{\perp}^2}  \frac{v_{\parallel} }{c}  Z_{\s} e  \bessel  \zeta,
	\nonumber \\
	\hat{M}_{A_{\parallel}}^{\dag} [\bp; \zeta] & = \left[1 + \frac{4 \pi}{ k_{\perp}^2  c^2} \sum_{\s} \frac{ (Z_{\s} e)^2 n_{\s}}{m_{\s}} \Gamma_{0, \s} \right] \zeta,
	\nonumber \\
	\hat{N}_{g,\s}^{\dag} [\bp; \sigma] & = 8 \pi \frac{J_{1,\s} }{a_{\s} } \mu_{\s}  \sigma ,
	\nonumber \\
	\hat{N}_{\phi}^{\dag} [\bp; \sigma] &=  \left[ 4 \pi \sum_{\s} \frac{Z_{\s} e n_{\s}}{B_0} \Gamma_{1, \s} \right] \sigma,
	\nonumber \\
	\hat{N}_{B_{\parallel}}^{\dag} [\bp; \sigma] &=  \left[1+ 16 \pi \sum_{\s} \frac{ n_{\s} T_{\s}}{B_0^2} \Gamma_{2, \s} \right] \sigma,
	\nonumber \\	
	\hat{C}_{\s}^{\dag} [\bp; \tildlam] & = \hat{C}_{\s} [\bp; \tildlam], 
\end{align}

\noindent
where we have defined
\begin{equation}
	\hat{S} [\bp; \tildlam] = v_{\parallel} \bdotgradzun \pdv{\tildlam}{z} - i \omega_{d,\s} \tildlam - i \omega_{*,\s} \tildlam.
\end{equation}

\section{Simplifying Adjoint Equations} \label{sec:Simplifying_adjoint_eqns}
Consider taking the following moments of \eqref{eq:lam_symbolic_eq}:
\begin{align}
	\left\langle \frac{Z_{\s} e}{T_{\s}} \bessel \maxwellian, \cdot \right\rangle_{v,\s} ,
	\quad
	\left\langle \frac{Z_{\s} e}{T_{\s}} \bessel \maxwellian \frac{v_{\parallel}}{c} , \cdot \right\rangle_{v, \s} ,
	\quad
	\left\langle 2 \frac{J_{1,\s}}{a_{\s}} \maxwellian \frac{\mu_{\s}}{T_{\s}} , \cdot \right\rangle_{v, \s} ,
\end{align}

\noindent
giving 
\begin{align}
	0 = \; & \sum_{\s} \frac{2 \pi B_0}{m_{\s}}  \int \rmd v_{\parallel} \int \rmd \mu_{\s}  \; \alpha_{\s} (z, v_{\parallel}, \mu_{\s} ) \maxwellian \left\{ \gamma^* \tildlam + v_{\parallel} \bdotgradzun \pdv{\tildlam}{z} \right. 
	\\ & 
	\left. - \frac{\mu_{\s}}{m_{\s}} \bdotgradzun \pdv{B_0}{z} \pdv{\tildlam}{v_{\parallel}} - i \omega_{d, \s} \tildlam + Z_{\s} e \bessel \xi - \frac{4 \pi}{k_{\perp}^2} Z_{\s} e \bessel \frac{v_{\parallel}}{c}  \zeta + 8 \pi \frac{J_{1,\s}}{a_{\s}} \mu_{\s} \sigma \right\}
	\label{eq:moments}
\end{align}

\noindent
where $\alpha_{\s}$ can take the forms:
\begin{align}
	\alpha_{\s} = \left\{
	\!\begin{array}{ll}
            \displaystyle
		\frac{Z_{\s} e}{T_{\s}} \bessel \\
            \displaystyle
		\frac{Z_{\s} e}{T_{\s}} \bessel \frac{v_{\parallel}}{c} \\
		\displaystyle 2 \frac{J_{1,\s}}{a_{\s}} \frac{\mu_{\s}}{T_{\s}} 
	\end{array}
	\right. .
\end{align}

\noindent 
We now identify different terms in \eqref{eq:moments} for each $\alpha_{\s}$ which are odd in $v_{\parallel}$ so evaluate to zero when integrated over the domain $\{-\infty, \infty \}$.:
\begin{align}
	0 = \; & \sum_{\s} \frac{2 \pi B_0} {m_{\s}} \int \rmd^2 v \frac{Z_{\s} e}{T_{\s}} \maxwellian \Bigg\{ \gamma^* \bessel \; \tildlam  + v_{\parallel} \bdotgradzun \bessel \pdv{\tildlam}{z} - i \omega_{d, \s} \bessel \tildlam 
	\nonumber \\
	& - \frac{\mu_{\s}}{m_{\s}} \bdotgradzun \pdv{B_0}{z} \pdv{\tildlam}{v_{\parallel}} + Z_{\s} e \bessel^2 \xi
	+ \underbrace{\frac{4 \pi}{k_{\perp}^2} Z_{\s} e \bessel^2 \frac{v_{\parallel}}{c}  \zeta }_{\text{odd in $v_{\parallel}$}} +  8 \pi \frac{J_{1,\s} \bessel }{a_{\s}} \mu_{\s} \sigma \Bigg\} ,
	\label{eq:first_moment}
\end{align}

\begin{align}
	0 = \; & \sum_{\s} \frac{2 \pi B_0} {m_{\s}} \int \rmd^2 v \frac{Z_{\s} e}{T_{\s}} \maxwellian \Bigg\{ \gamma^* \bessel \frac{v_{\parallel}}{c} \tildlam  + \frac{v_{\parallel}^2 }{c} \bdotgradzun \bessel \pdv{\tildlam}{z}
	\nonumber \\
	& - \frac{\mu_{\s}}{m_{\s}} \frac{v_{\parallel}}{c} \bdotgradzun \pdv{B_0}{z} \pdv{\tildlam}{v_{\parallel}} - i \omega_{d, \s} \frac{v_{\parallel}}{c} \bessel \tildlam 
	\nonumber \\
	&+ \underbrace{Z_{\s} e \bessel^2 \frac{v_{\parallel}}{c} \xi}_{\text{odd in $v_{\parallel}$}}
	+\frac{4 \pi}{k_{\perp}^2} Z_{\s} e \bessel^2 \left( \frac{v_{\parallel}}{c} \right)^2  \zeta  + \underbrace{ 8 \pi \frac{J_{1,\s} \bessel }{a_{\s}} \frac{v_{\parallel}}{c} \mu_{\s} \sigma}_{\text{odd in $v_{\parallel}$}} \Bigg\}  ,
	\label{eq:second_moment}
	\\
	0 = \; & 2 \sum_{\s} \frac{2 \pi B_0} {m_{\s}} \int \rmd^2 v  \maxwellian \Bigg\{ \frac{J_{1,\s}}{a_{\s}} \frac{\mu_{\s}}{T_{\s}} \gamma^* \; \tildlam  + v_{\parallel} \frac{J_{1,\s}}{a_{\s}} \frac{\mu_{\s}}{T_{\s}}  \bdotgradzun \pdv{\tildlam}{z} 
	\nonumber \\
	& -\frac{\mu_{\s}}{m_{\s}} \bdotgradzun \pdv{B_0}{z} \frac{J_{1,\s}}{a_{\s}} \frac{\mu_{\s}}{T_{\s}} \pdv{\tildlam}{v_{\parallel}} - i \omega_{d, \s} \frac{J_{1,\s}}{a_{\s}} \frac{\mu_{\s}}{T_{\s}} \tildlam + \frac{Z_{\s} e \mu_{\s}}{T_{\s}} \frac{\bessel \; J_{1,\s}}{a_{\s}} \xi
	\nonumber \\
	&
	+ \underbrace{ \frac{4 \pi}{k_{\perp}^2} Z_{\s} e \frac{\bessel \; J_{1,\s}}{a_{\s}} \frac{v_{\parallel}}{c} \frac{\mu_{\s}}{T_{\s}} \zeta }_{\text{odd in $v_{\parallel}$}} + 8 \pi \left(\frac{J_{1,\s} }{a_{\s}} \right)^2 \frac{\mu_{\s}^2 }{T_{\s}} \sigma \Bigg\} .
	\label{eq:third_moment}
\end{align}

\noindent
We can then perform integration by parts on the remaining $v_{\parallel}$ derivative, using velocity-independence of the fields. Using equations \eqref{eq:xi_symbolic_eq}-\eqref{eq:sigma_symbolic_eq} it is then possible to simplify the above equations to produce the results \eqref{eq:final_xi_eqn_EM}-\eqref{eq:final_sigma_eqn_EM}.

\section{Integral Weights} \label{appendix:mu_integral_weights}

\verb|stella| computes its $\tilde{\mu}_{\s}$ grid points according to Gauss-Laguerre quadrature such that the $\tilde{\mu}_{\s}$-integration of a variable, $f$, may be approximated numerically as 

\begin{equation}
	B \int_0^{\infty} \rmd \mu_{\s}  f(\mu_{\s}) \approx \sum_i^{N_{\mu_{\s}}} w_i e^{\hat{\mu}_{\s}} \frac{B}{B_0} f(\hat{\mu}_{\s}) ,
\end{equation}

where $N_{\mu_{\s}}$ is the number of grid points in $\mu_{\s}$, and the definition $\hat{\mu}_{\s} = \mu_{\s} B_0$ is made, with $B_0$ relating to the minimum of the normalised magnetic field. This $B_0$ acts as a scaling factor that acts on the upper-limit of the $\mu_{\s}$ grid such that the grid may be defined solely based on the number of grid points whilst still covering the necessary domain as $\hat{\mu}_{\s}$ is an independent coordinate that is independent of $\bp$. Previously, the importance of the role of the Jacobians was mentioned with a stress that certain variables are being considered as fixed, dummy variables, whilst others are dependent on the geometric inputs, $\bp$. We take the $\hat{\mu}_{\s}$ variable to be the dummy variable that is kept fixed in the perpendicular velocity integral such that the weights $w_i = B / B_0$ are the only factors to be perturbed in the integrals and, in a similar manner, the fixed variable, $\hat{\mu}_{\s}$, is weighted by a varying factor, $B_0$, when it appears in the equations. It should be noted that since the  $\hat{\mu}_{\s}$ grid is calculated independently of the geometry, it is the same for each perturbed value of $\bp$, and as such the $\hat{\mu}_{\s}$ grids will always align even when multiplying terms that are evaluated at different $\bp$'s, such as when coefficients are perturbed but are multiplied by $\gspec (\bp_0)$.
\section{Geometry Implementation} \label{sec:geometry}
The code \verb|stella| has an input option to use the Miller parametrisation of a flux surface; it takes a set of input variables to describe the local geometry of a specified flux surface along with the two adjacent flux surfaces on either side. We take  \eqref{eq:r_coordinate} in the form 

\begin{align}
	R(r, \theta) = R_0 (r) + r \cos [ \theta + \sin(\theta) \delta(r) ] ,
\end{align}
\noindent
with the triangularity redefined as $\delta (r) \coloneqq \arcsin [\bar{\delta}(r)]$. We now consider Taylor expanding in $r$ about $r =  r_{\psi_0}$

\begin{align}
	R_0(r) = \; & R_{0} ( r_{\psi_0}) + \left. \der{R_0}{r} \right|_{ r_{\psi_0}} (r -  r_{\psi_0} ) + ... \approx R_{\psi_0} + \Delta_{\psi_0} \rmd r + \mathcal{O} (\rmd r^2) , 
	\label{eq:major_rad_expansion}
\end{align}       

\noindent
with $\rmd r = r - r_{\psi_0}$, $R_{\psi_0} = R_0( r_{\psi_0})$, and $\Delta_{\psi_0} = \left. \der{R_0}{r}\right|_{ r_{\psi_0}}$. Similarly, 

\begin{align}
	\delta(r) = \; & \delta ( r_{\psi_0}) + \left. \der{\delta}{r} \right|_{ r_{\psi_0}} (r - r_{\psi_0}) + ... \approx \delta_{\psi_0} + \delta'_{\psi_0} \rmd r + \mathcal {O} (\rmd r^2), 
	\label{eq:triangularity_expansion}
\end{align}
\noindent
with $\delta_{\psi_0} = \delta ( r_{\psi_0})$, and $\delta'_{\psi_0} = \left. \rmd\delta / \rmd r \right|_{ r_{\psi_0}}$. Combining \eqref{eq:major_rad_expansion} and \eqref{eq:triangularity_expansion} gives

\begin{align}
	R(r,\theta) \approx &  R_{\psi_0} + r_{\psi_0} \cos[ \theta + \sin(\theta) \delta_{\psi_0} ] 
	\nonumber \\ 
	&+ \left\{ \Delta_{\psi_0} + \cos[\theta + \sin(\theta) \delta_{\psi_0} ] - r_{\psi_0} \sin[ \theta + \sin (\theta) \delta_{\psi_0} ] \sin(\theta)  \delta'_{\psi_0} \right\} \rmd r ,
\end{align}

\noindent
such that the above definition holds on any given flux surface, $\psi_0$. Equivalently, \eqref{eq:z_coordinate} can also be expanded about $r= r_{\psi_0}$ by first expanding the elongation

\begin{align}
	\kappa (r) = \kappa ( r_{\psi_0}) + \left. \der{\kappa}{r} \right|_{ r_{\psi_0}} (r - r_{\psi_0}) + ... \approx \kappa_{\psi_0} + \kappa'_{\psi_0} \rmd r + \mathcal{O} (\rmd r^2) ,
\end{align}

\noindent
with $\kappa_{\psi_0} = \kappa( r_{\psi_0})$, and $\kappa'_{\psi_0} = \left.\rmd \kappa / \rmd r \right|_{ r_{\psi_0}}$ to then write

\begin{equation}
	\begin{split}
		Z(r,\theta) \approx \; & r_{\psi_0} \kappa_{\psi_0} \sin(\theta) + \left[ \kappa_{\psi_0} + r_{\psi_0} \kappa'_{\psi_0} \right] \sin(\theta) \; \rmd r .
	\end{split}
\end{equation}

\noindent
These functions are used to describe the geometry of the flux surface of interest and the two adjacent flux surfaces by setting $\rmd r = \{0, \pm \Lambda \}$, with $\Lambda \ll 1$ a constant, in order to evaluate their radial derivatives. These quantities are then used to compute the Jacobian, magnetic field strength and configuration along with other functions defined on the flux surface.

\bibliography{jpp-instructions.bib}
\bibliographystyle{jpp}

\end{document}